\documentclass[onecolumn, a4size, 11pt,final]{IEEEtran}
\usepackage{amsmath}
\usepackage{amssymb}
\usepackage{amsfonts}
\usepackage[final]{graphicx}
\usepackage{ctable}
\usepackage{algorithm}
\usepackage{algorithmic}
\usepackage{multirow}
 \usepackage{subfigure}

 \title{Dynamic Base Station Operation in Large-Scale Green Cellular Networks}
 \author{Yue Ling Che,  Lingjie Duan, and Rui Zhang
 \thanks{Y.~L.~Che and L.~Duan are with the Engineering Systems and Design
 Pillar, Singapore University of Technology and Design (e-mail:
 yueling\_che@sutd.edu.sg;  lingjie\_duan@sutd.edu.sg).}
 \thanks{R.~Zhang is with the Department of Electrical and Computer Engineering,
 National University of Singapore (e-mail: elezhang@nus.edu.sg). He is also with
 the Institute for Infocomm Research, A*STAR, Singapore.}
 }

\begin{document}
\maketitle
\thispagestyle{empty}

\begin{abstract}
 In this paper, to minimize the on-grid energy cost in  a  large-scale green cellular network,  we jointly design the  optimal BS on/off operation policy 
and the on-grid energy purchase policy from a network-level perspective.   We consider  the BSs are aggregated as a microgrid with hybrid energy supplies 
and an  associated central energy storage (CES), which can store the harvested renewable energy and the purchased on-grid energy over time.
Due to the fluctuations of the on-grid energy prices, the harvested renewable energy, and the network traffic loads over time, as well as the BS coordination 
to hand over the traffic offloaded  from the inactive BSs to the active BSs, it is generally NP-hard to find a network-level optimal adaptation policy 
that can  minimize the on-grid energy cost over a long-term and yet assures the downlink transmission quality at the same time.
 Aiming at the network-level dynamic system  design, we jointly apply stochastic geometry (Geo) for large-scale green cellular network analysis 
 and  dynamic programming (DP) for adaptive BS on/off operation design and on-grid energy purchase design, and thus propose a new \emph{Geo-DP design approach}.
 By this approach, we  obtain the optimal BS on/off policy, which shows that the optimal BSs' active operation probability in each horizon is just sufficient 
 to  assure the required downlink transmission quality with time-varying load in the large-scale cellular network.
 However, due to the curse of dimensionality of the DP, it is of high complexity to obtain the optimal on-grid energy purchase policy. 
 We thus propose a suboptimal on-grid energy purchase policy with low-complexity, where the low-price on-grid energy is over-purchased 
 in the current horizon only when the current storage level and the future renewable energy level are both low.   
 Simulation results show that the  suboptimal on-grid energy purchase can achieve near-optimal performance.
 We also compare the proposed   policy with the existing schemes, to show  that our proposed policy can more efficiently save the on-grid energy cost over time.

\end{abstract}
 

\setlength{\baselineskip}{1.3\baselineskip}
\newtheorem{definition}{\underline{Definition}}[section]
\newtheorem{fact}{Fact}
\newtheorem{assumption}{Assumption}
\newtheorem{observation}{\underline{Observation}}
\newtheorem{theorem}{\underline{Theorem}}[section]
\newtheorem{lemma}{\underline{Lemma}}[section]
\newtheorem{corollary}{\underline{Corollary}}[section]
\newtheorem{proposition}{\underline{Proposition}}[section]
\newtheorem{example}{\underline{Example}}[section]
\newtheorem{remark}{\underline{Remark}}[section]
\newcommand{\mv}[1]{\mbox{\boldmath{$ #1 $}}}
\newtheorem{property}{\underline{Property}}[section]

\section{Introduction}
The dramatically increased mobile traffic  can easily lead to  an unacceptable total energy consumption level in the future cellular network, 
and energy-efficient cellular network design has become one of the critical issues for developing the 5G wireless communication systems \cite{E.A_Survey}. 
Due to the resulting high energy cost and the $\textrm{CO}_2$ emissions,  it is important to improve the cellular network energy efficiency (EE) 
by greening the cellular network  \cite{J.Wu_book}.
It   has been noted  that  the operations of base stations (BSs) contribute to most of the  energy consumptions in the cellular networks, 
which are estimated as  60-80 percent of the total network energy consumptions \cite{Marsan_ICC}. Hence, reducing the BSs' energy consumptions 
is crucial for developing energy-efficient or green cellular networks.

In traditional cellular networks, BSs are usually densely deployed  to meet  the peak-hour traffic load, which  causes unnecessary energy consumption 
for many lightly-loaded BSs during the low traffic load period (e.g., early morning). It is envisioned that in the future energy-efficient 
5G wireless communication system, the cellular network should dynamically shut down some BSs or  adapt the BS transmissions according to the time-varying  traffic  load. 
Although appealing,   due to   the significant spatial and temporal fluctuations of the traffic load in the cellular networks,    
optimal BS on/off operation  design is  a   challenging problem \cite{Luis.Magz}.
For example,  the authors  in \cite{Meo.ICC_workshop} studied   optimal energy saving for the cellular networks   by reducing the number of active cells,  
where  simplified cellular network model  with hexagonal cells and uniformly distributed traffic load is assumed.
In \cite{Niu.Mobih}, the authors proposed adaptive cell zooming scheme  according to the traffic load fluctuations, but without considering downlink transmission quality explicitly. 
In \cite{B.K.TWC.13},  by studying the impact of each BS's active operation on the network performance, 
the authors proposed a suboptimal BS on/off scheme that can be implemented in a distributed manner to save the BSs' energy consumption. 
However,  the impact of traffic fluctuation on the BS's on/off decisions as well as the downlink transmission quality were  not  explicitly  considered.
The issue of downlink transmission quality was considered in \cite{Luo.TWC.13}, where the BS's transmit power and cell range adaptations 
to the traffic load fluctuation were proposed to improve the EE of the cellular network, but \cite{Luo.TWC.13} only considered a single-cell scenario.
 In \cite{Xu.JSAC.15}, an energy group-buying scheme  was proposed to let     the network operators  make the day-ahead and real-time energy purchase  as a single group, 
 where the BSs of the network operators share the wireless traffic to maximally turn lightly-loaded BSs into sleep mode.

To more effectively reduce the carbon footprint and improve the EE,  the energy harvesting (EH) enabled BSs, 
which are able to harvest   clean and free renewable energy (e.g., solar  and wind energy)  from the surrounding 
environment for utilization, have been developed for the green cellular networks recently. 
For example, in \cite{Ho.TSP.12}, by assuming a renewable energy storage, the authors applied the dynamic programming (DP) technique 
to study the energy allocations for a point-to-point transmission, to adapt to the time-selective channel fading.    
In \cite{Dhillon.TWC.14}, by using a birth-death process to model the harvested renewable energy level at each  BS for  a large-scale 
$K$-tier green cellular network,  the authors characterized an energy availability region of the  BSs  based on tools from stochastic geometry \cite{Stoyan.SG.95}. 
Various issues of the EH-enabled wireless communications are also discussed in \cite{Gunduz.ComMag.14}-\cite{Xu.ComMag.15}. 
However,  it is noted that the available renewable energy in nature is limited and  intermittent, and thus cannot always guarantee sufficient energy supply to power up the BSs.

By combining the merits of the reliable on-grid energy and the cheap renewable energy, hybrid energy supplied BS design 
has become a promising method to power up the BSs \cite{Han.TWC.13}. 
One of the main design issues  is to properly exploit the available renewable energy to  minimize the on-grid energy cost or maximize the EE of the system. 
We note that some initial work   has been proposed to address this issues via,  
e.g.,   the optimal packet scheduling  \cite{G.Yu.TWC.14},  the optimal BS  resource allocation \cite{Ng.TWC.13}, 
and    low-complexity online algorithm design   based on the Lyapunov optimization \cite{Mao.JSAC.15}.
In \cite{Xu.TVT.14}-\cite{Guo.TCM.14},  to mitigate the fluctuations of the renewable energy harvested by each BS, 
the authors also considered energy cooperation  and sharing between the BSs. 
Moreover, by applying the DP technique to adapt to the fluctuations of both the traffic loads and the harvested renewable energy,  
hybrid energy supplied BS on/off operation design   has also been studied in the literature. 
For example, in \cite{Zhou.ICTC.13}, by assuming each BS in the network is supplied by either the renewable energy or the on-grid energy,  
the authors studied the BS sleep control problem to save the energy consumptions of all the BSs. 
In \cite{Niu.TCOM.14}, by assuming each BS is jointly supplied by renewable energy and on-grid energy, the authors  studied the 
joint  optimization of the BS on/off operation and resource allocation  under a user blocking probability constraint at each BS. 
Due to the traffic offloaded from the inactive BSs to the active BSs,   all the BSs in the network are essentially coordinated to support 
to all the users in both \cite{Zhou.ICTC.13} and \cite{Niu.TCOM.14}. However,  due to the dimensionality curse of the DP \cite{DP}, 
it is generally \emph{NP-hard} to conduct a network-level BS coordination to decide their optimal on/off status. 
As a result, in   \cite{Zhou.ICTC.13} and \cite{Niu.TCOM.14}, the authors proposed a cluster-based BS coordination scheme 
to decide the BSs' on/off status in a large-scale network, where the BSs form different clusters, and the BSs in each cluster 
apply the DP technique to decide their on/off status. However, such a cluster-based BS coordination scheme cannot achieve 
network-level optimal BS on/off decisions in general.

In this paper, we consider a large-scale cellular network, where all the BSs are aggregated as a microgrid with hybrid energy supplies 
and  an associated central energy storage (CES). The microgrid works in island mode if it has enough local supply from the renewable farm; 
otherwise, it connects to the main grid for purchasing the on-grid energy \cite{Fang.Surveys.12}.
 By adapting to the fluctuations of the harvested renewable energy, the network traffic loads, as well as the on-grid energy prices over time, 
 the microgrid applies the DP technique to decide the amount of on-grid energy to purchase to satisfy all the BSs' energy demand in the network, 
 as well as the probability that each BS stays active, which is referred to as  the BSs' active operation probability, at each  time, 
 so as to minimize the total on-grid energy cost over time in the network. 
 Unlike \cite{Zhou.ICTC.13} and \cite{Niu.TCOM.14}, we aim at a network-level optimal BS on/off operation policy design, 
 such that the on-grid energy cost in the cellular network can be efficiently saved.

The key contributions of this paper are summarized as follows:
\begin{itemize}
 \item \emph{Novel large-scale network model with hybrid-energy supplied BSs: }
Based on  Poisson point processes (PPPs), we first   model the large-scale network by jointly considering the mobility of mobile terminals (MTs),   
traffic offloading from an inactive BS to its neighboring active BSs, as well as BS frequency reuse for reducing the downlink network interference level.   
Since all the BSs are supplied by the hybrid energy stored in the CES, we then  model the hybrid energy management at the CES, 
where the on-grid energy is purchased with time-varying price  to ensure that the energy demands of all the BSs in the network can be satisfied.

 \item \emph{New Geo-DP approach for dynamic network optimization:}
 To pursue a network-level design, we jointly apply stochastic geometry (Geo) for large-scale green cellular network analysis 
 and  the DP technique for adaptive BS on/off operation  and on-grid energy purchase, and thus propose a new \emph{Geo-DP approach}. 
By this approach,  the microgrid  centrally decides the  optimal BSs' active operation probability  and the purchased   on-grid energy   in each time horizon, 
 so as to minimize the total on-grid energy cost over all time horizons, under  the BSs' downlink transmission quality constraint and the BSs' energy demand constraint.

 \item \emph{Network-level optimal policy design:}
  We aim at optimal offline  policy design, which  updates the BSs' on/off status and on-grid energy purchase decisions in a predictable future period. 
 By studying the impact of the BSs' active operation probability on the storage level  at the CES, we   find an optimal BS on/off policy, 
 which shows that the optimal BSs' active operation probability in each horizon is just sufficient to  assure the required downlink transmission quality 
 with time-varying load in the large-scale cellular network.
 However, due to the curse of dimensionality of the DP, it is of high complexity to obtain the optimal on-grid energy purchase policy.
 We thus propose a suboptimal on-grid energy purchase policy with low-complexity, where the low-price on-grid energy is over-purchased 
 in the current horizon only when the current storage level and the future renewable energy level are both low.

 \item \emph{Efficient network-level BS coordination to minimize the on-grid energy cost:}  By conducting extensive simulations in Section VI, 
 we show that the proposed suboptimal on-grid energy purchase policy can achieve near-optimal performance. 
 We also show that   our proposed policy is robust to the prediction errors of the on-grid energy prices, the harvested renewable energy, 
 and the network traffic loads, and thus can be properly implemented in practice. Moreover, we conduct a large-scale network simulation 
 to compare  our proposed policy with the existing schemes, and   show that our proposed policy can more efficiently save the on-grid energy cost in the network.

\end{itemize}

As a powerful tool, stochastic geometry has been widely applied to model the EH enabled wireless communication networks. 
 For example, besides  \cite{Dhillon.TWC.14}, we also noticed that the EH-based ad hoc networks, cognitive radio networks, 
 cooperative communication networks have been studied in \cite{Huang.IT.13}, \cite{Lee.TWC.2013}, and \cite{Khan.ICC}, respectively, 
 by adopting the PPP-based network modeling.
 In our previous work \cite{Che.JSAC.15}, we also applied tools from stochastic geometry to analyze the spatial throughput of a wireless powered  communication network, 
 where the users exploit  their harvested radio frequency (RF) energy to power up their communications.
Although the network-level system performance can be  characterized in these existing studies,
the adaptation   to the fluctuations of the network parameters (e.g., the harvested renewable energy, the network traffic loads, and the on-grid energy prices) 
has not been properly addressed. To our best knowledge, our  newly proposed Geo-DP approach that can lead to a network-level optimal adaptation policy design 
has not been studied in the literature.

\section{Network Operation Model} \label{section: system_operation}
\begin{figure}
\setlength{\abovecaptionskip}{-0.04in}
\centering
\DeclareGraphicsExtensions{.eps,.mps,.pdf,.jpg,.png}
\DeclareGraphicsRule{*}{eps}{*}{}
\includegraphics[angle=0, width=0.8\textwidth]{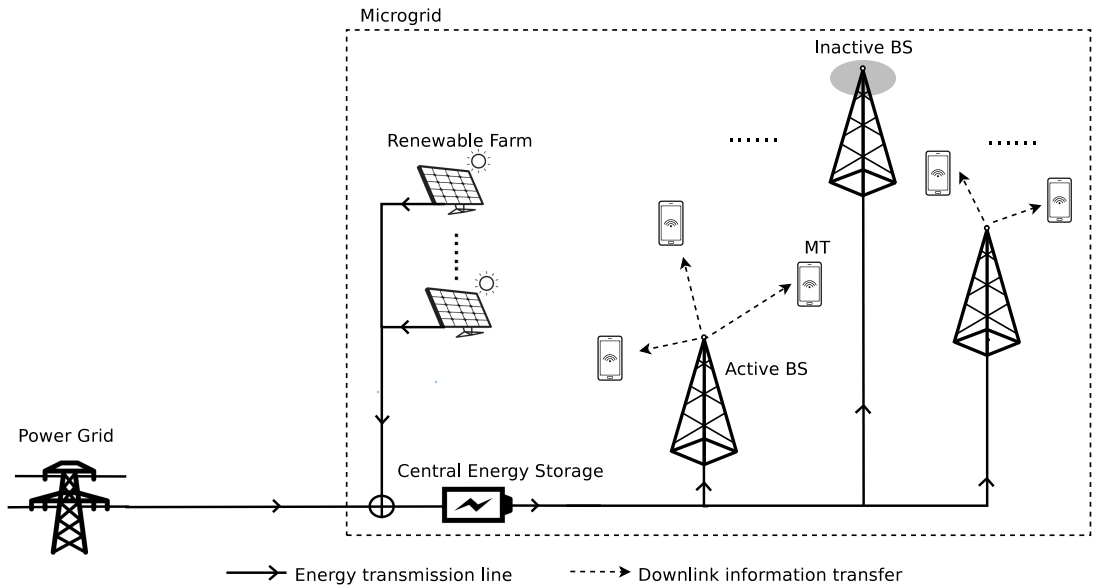}
\caption{ Illustration of the cellular network supplied by both renewable energy and on-grid energy.}
\label{fig: network_model}
\vspace{-0.1in}
\end{figure}
We consider a   large-scale    cellular  network, where the traffic loads, the on-grid energy prices,  the available renewable energy,  as well as
 the  wireless fading channels from each BS to the MTs  are all varying over time in the network.  The BSs are able to switch between active  (i.e., on status)
 and inactive (i.e., off status) operation modes  to save the total on-grid energy cost over time.
 Supported by smart grid development, as illustrated in Fig.~\ref{fig: network_model}, we aggregate all the BSs as a microgrid with hybrid energy supplies 
 and an associated CES   to defend against the supply/load variation \cite{Fang.Surveys.12}. For efficient energy management, 
 the cellular network deploys the   CES in the microgrid, which connects to both the local renewable farm (e.g., wind turbines or solar panels) and the power grid, 
 and thus can  store the harvested  renewable energy (e.g., wind or solar energy) from the environment  and
 the purchased  on-grid energy from the power grid at the same time{\footnote{A CES is easier and cheaper to manage as compared to distributed storage at each BS. 
 One can also view this CES as a collection of distributed storage with perfect energy exchange links.}}.
 All the BSs are connected with the CES and are powered up by its stored hybrid energy.
 Similar to \cite{Luo.TWC.13}, it is assumed that the CES is able to charge and discharge at the same time.
In the following subsections, we first model each BS's operations over time, and  then  present the   large-scale cellular network model based on homogeneous PPPs.
The hybrid energy management model at the CES will be elaborated later in Section III.

\subsection{Time Scales for Network  Operations} \label{section: horizon_on_off}

\begin{figure}
\setlength{\abovecaptionskip}{-0.04in}
\centering
\DeclareGraphicsExtensions{.eps,.mps,.pdf,.jpg,.png}
\DeclareGraphicsRule{*}{eps}{*}{}
\includegraphics[angle=0, width=0.6\textwidth]{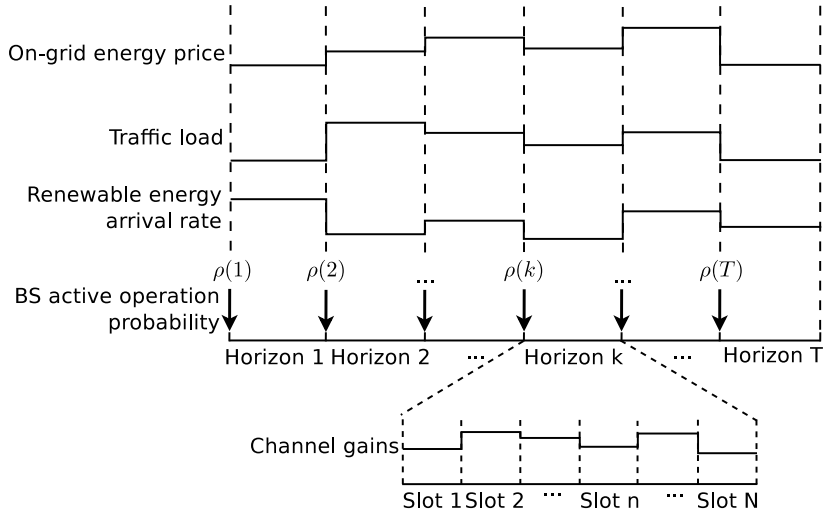}
\caption{ Dynamic network operations over two different time scales: horizons and slots. }
\label{fig: horizon_model}
\vspace{-0.1in}
\end{figure}

 As compared to the variations of the downlink wireless fading channels from each BS to the MTs,   the renewable energy arrival rates, the network traffic loads,   
 as well as  the on-grid energy prices  are  all varying  slowly over time  \cite{Dhillon.TWC.14}.
We thus consider the network operates over two different time scales \cite{Li.ComMag.15}: one is the long time scale and is referred to as \emph{horizons}, 
and the other is the short time scale and is referred to as \emph{slots}, as shown in Fig.~\ref{fig: horizon_model}. 
We define   a horizon as a reasonably large time period where   the renewable energy arrival rate, the network traffic load, and the on-grid energy price all remain unchanged. 
Each horizon consists of $N$  slots, and the wireless fading channels vary over slots.   
We focus on  in total $T$ horizons, $1\leq T <\infty$, and assume each horizon has a fixed duration $\tau>0$. 
For simplicity, we assume $\tau=1$ hour in the sequel, and thus interchangeably use Wh and W to measure the energy consumption in each horizon.

By adapting to the variations of the on-grid energy prices,   the renewable energy arrival rates, the network traffic loads, as well as the wireless fading channels,
  the microgrid centrally decides the amount of  on-grid energy to purchase and the  active operation probability of the BSs in the network for  each horizon $t\in\{1,...,T\}$,
which are denoted by $G(t)>0$ and   $\rho(t)$, respectively, so as  to minimize the total on-grid energy cost over all $T$ horizons.
Similar to  \cite{Gunduz.ComMag.14}, we consider the optimal offline policy design, and 
  assume the fluctuations of the renewable energy arrival rates, the network traffic loads, as well as the on-grid energy prices in all $T$ horizons can be accurately predicted \cite{Chiang.JSAC.12}.
  We will show that even given reasonable prediction errors, our approach works quite well in Section VI by simulation.
In each horizon $t$, based on   $\rho(t)$ determined by the microgrid, each BS independently decides to be active or inactive. If a BS decides to be active, it
transmits to its associated MTs  with a fixed power level \cite{LTE}, denoted by $P_B>0$. If a BS decides to be inactive, it keeps silent  and its originally associated MTs are handed over to its neighboring active BSs.

\subsection{PPP-based  Network Model} \label{section: network_operation}
This subsection models the large-scale cellular network based on stochastic geometry.
We assume  the BSs are deployed randomly according to a homogeneous PPP, denoted by $\Phi(\lambda_B)$, of  BS density $\lambda_B>0$.
For a given active operation probability $\rho(t)$ in horizon $t$,  due to the independent on/off decisions at the BSs,
 the point process formed by the active BSs in each horizon $t\in \{1,...,T\}$ is still a homogeneous PPP, denoted by $\Phi(\lambda_B\rho(t))$, 
 with active BS density given by $\lambda_B\rho(t)$.
 We assume  the MTs are also distributed as a  homogeneous PPP, denoted by $\Phi(\lambda_m(t))$, of MT density $\lambda_m(t)>0$ in horizon $t$.
All $\Phi(\lambda_m(t))$'s,  $\forall t\in\{1,...,T\}$, and $\Phi(\lambda_B)$ are  mutually independent.
The MTs   independently move in the network over slots in each horizon based on the random walk model   \cite{Baccelli.NOW.I}.
Each MT is associated with its nearest active BS from  $\Phi(\lambda_B\rho(t))$ in  each slot of  horizon~$t$,   for achieving good communication quality.
When an active BS becomes inactive, its originally associated MTs   will be  transferred to their nearest BSs that are still active   as in \cite{Meo.ICC_workshop}-\cite{Niu.Mobih} and \cite{G.Yu.TWC.14}-\cite{Niu.TCOM.14}.
From \cite{Andrews.COM.11}, the resulting  coverage areas of the active BSs in each slot comprise a Voronoi tessellation on the plane $\mathbb{R}^2$.
We refer to   {traffic load} of an active BS as the number of MTs that are associated with it, i.e., those are located  within the  Voronoi cell of the active BS, as in \cite{Dhillon.TWC.14}.
The following proposition studies the average traffic load of an active BS.
\begin{proposition} \label{proposition: traffic_load}
Given the BSs' active operation probability $\rho(t)$ in horizon $t$, $\forall t \in \{1,...,T\}$,     the average  traffic load for an active BS   from $\Phi(\lambda_B\rho(t))$ is  given by $ D(t)=\frac{\lambda_m(t)}{\lambda_B\rho(t)}$.
\end{proposition}

The proof of Proposition \ref{proposition: traffic_load} is obtained by noticing that $D(t)$ equals   the product of the  MT density $\lambda_m(t)$ 
and the average coverage area of each BS, where the latter term is  $\frac{1}{\lambda_B\rho(t)}$ from Lemma 1 in \cite{Dhillon.TWC.14},  and thus is omitted here for brevity.
From Proposition~\ref{proposition: traffic_load}, the average  traffic load $D(t)$  at each BS is increasing over the MT density $\lambda_m(t)$, 
and decreasing over the BS density $\lambda_B$ as well as the BSs' active operation probability $\rho(t)$, as expected. 
Moreover, since the inactive BSs do not transmit information to the MTs and thus have zero traffic load,    
the average traffic load for each BS from  $\Phi(\lambda_B)$ is also obtained as $D(t)$ for each horizon $t\in \{1,...,T\}$.

\subsection{Frequency Reuse Model} \label{section: freq_reuse}
To effectively control    the potentially large downlink interference,  which is mainly caused by the nearby BSs that are operating over the same frequency band,  
this subsection presents    a \emph{random frequency reuse model}   to assign different frequency bands to the neighboring BSs.
Specifically, we assume the total bandwidth available for the network is $W$ Hz. To properly support the traffic load $D(t)$, we assume the total bandwidth is equally divided into 
$\delta(t)$ bands in each horizon $t$, and each BS is randomly assigned with one out of $\delta(t)$ bands for operation as in \cite{Andrews.COM.11}. 
We consider orthogonal multiple access for the MTs in the Voronoi cell of each active BS as in, e.g.,  \cite{Dhillon.TWC.14} and  \cite{Andrews.COM.11}, 
and each MT is assigned with a  channel with fixed bandwidth $B$ Hz for communication, where $0<B<W$. 
Since each active BS is only assigned with one band, by letting each BS's required total bandwidth to support its associated MTs 
be equal to the average bandwidth per band, i.e., $D(t)B=W/\delta(t)$,  we obtain
\begin{equation}
 \delta(t)=\frac{W}{B}\frac{1}{D(t)}=\frac{W}{B}\frac{\lambda_B \rho(t)}{\lambda_m(t)}, \label{eq: delta}
\end{equation}
by substituting $D(t)=\frac{\lambda_m(t)}{\lambda_B\rho(t)}$  from Proposition \ref{proposition: traffic_load}.
It is observed from (\ref{eq: delta}) that  the number of bands $\delta(t)$ decreases as the traffic load $D(t)$ increases, as expected.
For each horizon $t\in \{1,...,T\}$ with a given $\delta(t)$, due to the BSs' independent operation band selection as well as  the independent on/off decisions over horizons,  
the point process formed by the co-band active BSs   is still a homogeneous PPP,  denoted by  $\Phi(\lambda_B^c(t))$, with co-band active BS density $\lambda_B^c(t)=\lambda_B\rho(t)\frac{1}{\delta(t)}$. 
As a result, for each MT, it only   receives interference from all other active BSs that operate over the same band as its associated BS with probability $1/\delta(t)$.
In the next subsection, based on the PPP-based network model and the frequency reuse mode, we present the downlink transmission model from each BS to its associated MT.

 \subsection{Downlink Transmission Model} \label{section: transmission_model}
 We measure the downlink transmission quality between the BS and its associated MTs based on the received signal-to-interference-plus-noise-ratio (SINR) at the MTs.
We  assume  Rayleigh flat fading channels with path-loss as in  \cite{Dhillon.TWC.14},  \cite{Huang.IT.13}, \cite{Che.JSAC.15}, and  \cite{Andrews.COM.11}, 
where the channel gains vary over slots as  shown in Fig.~\ref{fig: horizon_model}.  Let $\Phi(\lambda_B)=\{X\}$, $X \in\mathbb{R}^2$, denote   the coordinates  of the  BSs.  
In slot $n\in \{1,...,N\}$ of horizon $t\in \{1,...,T\}$,  let $h_{X}(t, n)$ represent an  exponentially distributed random variable  with unit  mean 
to model Rayleigh fading   from BS $X$  to the origin $o=(0,0)$ in the plane $\mathbb{R}^2$. 
We assume $h_{X}(t, n)$'s are mutually independent over any time slot $n\in \{1,...,N\}$, any horizon $t\in \{1,...,T\}$, and any BS $X\in \Phi(\lambda_B)$.
Suppose a typical MT is located at the origin and is associated with the active BS $X\in \Phi(\lambda_B\rho(t))$ in slot $n$ of horizon $t$.
We express the desired signal strength received at the typical MT  as  $|X|^{-\alpha}h_X(t,n)$, where $|X|$  is the distance  from the associated active BS $X$  to the origin, 
$\alpha>2$ is the path-loss exponent. 
Under the random frequency reuse model, since the received interference at the typical MT is caused by the co-band BSs from $\Phi(\lambda_B^c(t))$, 
we  express the received interference at the typical MT    as $\sum_{Y\in \Phi(\lambda_B^c(t)), Y\neq X}P_B|Y|^{-\alpha}h_Y(t,n)$.  
Thus, the received  SINR  at the typical MT from BS $X$   in slot $n$ of horizon $t$, denoted by $\textrm{SINR}_o(t,n)$, is  obtained as
\begin{equation}
\textrm{SINR}_o(t,n)=\frac{P_B|X|^{-\alpha}h_X(t,n)}{\sum_{Y\in \Phi(\lambda_B^c(t)), Y\neq X}P_B|Y|^{-\alpha}h_Y(t,n)+\sigma^2}, ~~\forall t\in \{1,...,T\}, \forall n\in \{1,...,N\}, \label{eq: SINR_o_def}
\end{equation}
where $\sigma^2$ is the noise power at the typical MT. It is also noted that  $|X|$ in (\ref{eq: SINR_o_def}) also represents the minimum distance from the typical MT to
all the active BSs from $\Phi(\lambda_B\rho(t))$, regardless of their operating frequency bands, i.e., no other active BSs can be closer
than $|X|$. By using the null probability of a PPP, the probability density function (pdf) of $|X|$ in (\ref{eq: SINR_o_def}) is then obtained
 as  $ f_{|X|}(r)=2\pi \lambda_{B}\rho(t) r e^{-\lambda_{B}\rho(t) \pi r^2}$ for $r\geq0$ \cite{Andrews.COM.11}.
We say a downlink transmission is successful if the received SINR at the MT is no smaller than a predefined threshold $\beta>0$.

 \section{ Hybrid Energy Management Model}  \label{section: energy_model}
 This section  presents the hybrid energy management model at the CES, by modeling the   on-grid and renewable energy consumptions of all the BSs in the network. 
 In the following, we first model the energy consumption at  both active and inactive BSs. 
 We then detail the hybrid energy management at the CES so  as to meet the  energy demands of the BSs.

\subsection{ Energy Consumption at the BS} \label{section: energy_demand}
According to the BS energy consumption breakdown provided by the practical studies  in \cite{Earth}, 
various components of the BS, e.g., the  power amplifier,  base-band operation, and power supply for power grid and/or back-haul connections, 
all contribute   to the total BS energy consumption. It is also shown in \cite{Earth} that the BS energy consumption can be well approximated 
by a \emph{linear power model}, where  the total energy consumption of an active transmitting BS   is affine/linear to its traffic load, 
with the corresponding slope determined by the power amplifier efficiency.  
Such a linear power model has been widely adopted in the literature (see, e.g.,  \cite{Ho.TSP.12} and \cite{Niu.TCOM.14}). 
We thus also   adopt it to model the energy consumption for each BS  in this paper.

Specifically, for a given  horizon $t$, although the traffic load  of each active BS generally varies over space, 
by  applying our PPP-based network model  in Section \ref{section: system_operation}, we can obtain the average energy consumption of each active BS over space as
\begin{equation}
 P_{on}(t)=P_a+\frac{P_B}{\mu}D(t), t\in\{1,...,T\},\label{eq: P_on}
\end{equation}
where the average traffic load $D(t)$ of the active BS is given in Proposition \ref{proposition: traffic_load},  $\mu\in(0,1)$ is the power amplifier efficiency,  
and  $P_a>0$  represents  the active BS's average energy consumption for supporting its basic operations and is taken as a constant    \cite{Earth}.  
For an inactive BS in horizon $t$, since it keeps silent in the horizon, we model its average energy consumption as a constant, which is given by
\begin{equation}
 P_{off}(t)= P_s, t\in\{1,...,T\}, \label{eq: P_off}
\end{equation} 
where similar to $P_a$ in (\ref{eq: P_on}), the constant   $P_s>0$  represents the average energy consumption of an inactive BS for supplying its basic operations. 
We assume $P_a>P_s$ \cite{Earth}.
Based on the energy consumption model for each active or inactive BS, we present the hybrid energy management model at the CES to meet all the BSs' energy demand.

\subsection{Energy Supplies at the CES} \label{section: energy_supplies}
As shown in Fig.~\ref{fig: network_model}, the CES can store both the renewable energy that is harvested from the local renewable farm and the on-grid energy that
is purchased from the power grid.
In each horizon, the microgrid centrally decides the amount of  on-grid energy to purchase from the power grid and  the BSs' active operation probability $\rho(t)$, 
by jointly considering
the BSs' downlink transmission quality,  the energy demands of all the BSs in each horizon, and the hybrid energy storage level at the CES.
We focus on the average energy consumption of all the BSs normalized by  spatial area.
For a given active operation probability $\rho(t)$ in horizon $t$, by considering the BS density $\lambda_B$ and its average traffic load $D(t)$ in each horizon $t$, 
we can obtain the average energy consumption of all the BSs
over the BSs' on/off status  and the network area in each horizon as
\begin{align}
 E(t)&=\lambda_B \rho(t)P_{on}(t)+\lambda_B(1-\rho(t))P_{off}(t) \label{eq: E_t_def}\\
 &\overset{(a)}{=}\lambda_B \rho(t) P_{gap}+\lambda_B P_s+\frac{P_B\lambda_m(t)}{\mu}, t\in \{1,...,T\}. \label{eq: E_t}
\end{align}
 where equality $(a)$ follows by substituting   (\ref{eq: P_on}) and (\ref{eq: P_off}) into  (\ref{eq: E_t_def}), and $P_{gap}=P_a-P_s>0$. 
 It is easy to find that the unit of $E(t)$  is given by W$/\textrm{unit-area}$. 
 It is also observed from (\ref{eq: E_t}) that $E(t)$ increases monotonically over the BSs' active operation probability $\rho(t)$ in each horizon $t$.

 Denote the storage level of the CES at the beginning of horizon $t$ as $B(t)\geq 0$,   the amount of  on-grid energy to purchase in horizon $t$ as $G(t)$, 
 and the renewable energy arrival rate in horizon $t$ as $\lambda_e(t)$. All $B(t)$, $G(t)$, and $\lambda_e(t)$ have the same unit as $E(t)$, i.e., W$/\textrm{unit-area}$. 
 According to Fig.~\ref{fig: network_model}, to  satisfy the demanded energy of all the BSs in horizon $t$,  we obtain the following energy demand constraint:
 \begin{equation}
   B(t)+G(t)+\lambda_e(t)\geq E(t),  t\in \{1,...,T\}. \label{eq: energy_demand_constraint}
 \end{equation}
If the purchased on-grid energy $G(t)$  and/or harvested renewable energy $\lambda_e(t)$ cannot be completely consumed by the BSs in the current horizon, 
their residual portions are both stored in the CES for future use.
The dynamics of the storage level $B(t)$ is obtained as
\begin{equation}
 B(t+1)=\min\left(B(t)+ \lambda_e(t)+G(t)-E(t),C\right),  \label{eq: B_t}
\end{equation}
where $0<C<\infty$ is the given storage capacity of the CES. For simplicity, we assume  the initial storage level  at the beginning of horizon $t=1$ is  $B(1)=0$.

\section{Performance Metrics and Problem Formulation} \label{section: problem_formulation}
In this section, based on the network operation model and the hybrid energy management model 
in Sections~\ref{section: system_operation}~and~\ref{section: energy_model}, respectively, 
we propose  a new Geo-DP approach to study the cellular network and formulate the on-grid energy cost minimization 
problem from the network-level perspective. 
In particular, we first use stochastic geometry to characterize the downlink transmission performance from the BSs 
to their associated MTs. 
We then further apply DP to formulate the   on-grid energy cost minimization problem under the BSs' successful downlink transmission probability constraint.

\subsection{Successful Downlink Transmission Probability}
As discussed in Section \ref{section: system_operation},  if a BS is active in a particular horizon, it transmits to its associated MTs in each slot within this horizon, 
while the channel fadings   from the BSs to the MTs as well as the locations of the MTs vary over slots.
In this subsection, by applying tools from stochastic geometry, we   study  the active BSs' successful downlink transmission probability to their associated MTs in each slot. 
Specifically,  from Section \ref{section: network_operation}, due to the stationarity of the considered homogeneous PPPs for the active BSs and the MTs in each horizon 
as well as the independent moving of the MTs over slots,    we focus on a typical   MT   in each slot of horizon $t\in \{1,...,T\}$, 
which is assumed to be located at the origin of the plane $\mathbb{R}^2$, without loss of generality.   
Denote the successful downlink transmission probability for  the typical   MT   in slot $n$ of horizon $t$ as 
$\mathcal{P}_{suc}(t,n)$, $\forall t\in \{1,...,T\}$, $\forall n\in \{1,...,N\}$. 
According to Section \ref{section: transmission_model}, for a targeted SINR threshold $\beta$, we define  $\mathcal{P}_{suc}(t,n)$  as
\begin{equation}
 \mathcal{P}_{suc}(t,n)=\mathbb{P}\left(\textrm{SINR}_o(t, n) \geq \beta \right), \forall t\in \{1,...,T\}, \forall n\in \{1,...,N\}, \label{eq: P_suc}
\end{equation}
with  $\textrm{SINR}_o(t, n)$  for a typical MT given in (\ref{eq: SINR_o_def}).
By applying the probability generating functional (PGFL) of a PPP \cite{Stoyan.SG.95},  we explicitly express $\mathcal{P}_{suc}(t,n)$   in the following proposition.

\begin{proposition} \label{proposition: P_suc}
 In slot $n$ of horizon $t$, $\forall t\in \{1,...,T\}$, $\forall n\in \{1,...,N\}$, the successful downlink transmission probability for  the typical  MT  is
 \begin{equation}
   \mathcal{P}_{suc}(t,n)=\pi \lambda_B \rho(t) \int_0^{\infty} e^{-ax}e^{-bx^{\alpha/2}}\, dx, \label{eq: P_suc_general}
 \end{equation}
with $a=\pi \lambda_B \rho(t)+\pi\frac{\lambda_m(t)B}{W}v$, where $B$ and $W$ are the fixed channel bandwidth for each MT and the total  bandwidth available 
in the network, respectively, as introduced in Section \ref{section: freq_reuse},  and $v=\beta^{2/\alpha}\int_{\beta^{-2/\alpha}}^{\infty}\frac{1}{1+u^{\alpha/2}} \, d u$, and $b=\beta\sigma^2 /P_B$. When $\alpha=4$,
 (\ref{eq: P_suc_general}) admits a closed-form expression with
  \begin{equation}
 \mathcal{P}_{suc}(t,n)=\pi \lambda_B \rho(t)\sqrt{\frac{\pi P_B}{\beta \sigma^2}} \exp\left(\frac{\Upsilon(t)^2}{2} \right)Q(\Upsilon(t)), \label{eq: P_suc_alpha4}
\end{equation}
where  $\Upsilon(t)=\frac{\left(\pi \lambda_B \rho(t)+\pi\frac{\lambda_m(t)B}{W}v\right)\sqrt{P_B}}{\sqrt{2\beta \sigma^2}}$ with
$v= \sqrt{\beta}\left(\pi/2-\arctan\left( 1/\sqrt{\beta}\right)\right)$, and
$Q(x)=\frac{1}{\sqrt{2 \pi}}\int_x^{\infty}\exp\left(-\frac{u^2}{2} \right)\, du$ is the standard Gaussian tail probability.
\end{proposition}

Proposition \ref{proposition: P_suc} is proved using a method similar to that for proving Theorem 2 in \cite{Andrews.COM.11}, and thus is omitted for brevity. 
By using a similar method as in our previous work \cite{Che.JSAC.15}, one can also easily validate Proposition \ref{proposition: P_suc} by simulation.  
Moreover, from Proposition \ref{proposition: P_suc}, due to the independent moving  of the MTs as introduced in Section \ref{section: network_operation}, 
the expressions of $\mathcal{P}_{suc}(t,n)$ are observed to be identical for all slots within each horizon $t \in \{1,...,T]\}$. 
Thus, for notational simplicity, we omit the slot index $n$ and use $\mathcal{P}_{suc}(t)$ to represent the typical BS's   
successful downlink transmission probability in any slot of horizon $t$ in the sequel.

\subsection{Problem Formulation} \label{section: formulation}
Denote the on-grid energy price in each horizon $t$ as $a(t)$ (\$/W). The total on-grid energy cost across all the BSs  over all $T$ horizons is thus obtained 
as $\sum_{t=1}^T a(t)G(t)$, with unit $\$/\textrm{unit-area}$. 
To minimize the total on-grid energy cost under the energy demand constraint given in (\ref{eq: energy_demand_constraint}), 
the microgrid can take  advantage of the on-grid energy's price variations by purchasing more on-grid energy when its price is low and store it for future use, 
and purchasing less (or even zero) on-grid energy when its price is high. 
To  ensure the QoS for each BS's downlink transmissions, we  also apply a \emph{successful downlink transmission probability constraint} 
such that $\mathcal{P}_{suc}(t)\geq 1-\epsilon$  with $\epsilon\ll 1$ and $t\in \{1,..., T\}$  \cite{Andrews.COM.11}.
As a result, by jointly considering the energy demand constraint and the successful downlink transmission probability constraint, 
we optimize the BSs' active operation probability $\rho(t)$ and the purchased on-grid energy $G(t)$ in each horizon, 
and formulate the on-grid energy cost minimization problem as
 \begin{align}
\textrm{(P1)}:~~~\mathop{\mathrm{min.}}_{\rho(t),G(t)}&~~   \sum_{t=1}^{T} a(t)G(t) \nonumber \\
\mathrm{s.t.} & ~~  \mathcal{P}_{suc}(t)\geq 1-\epsilon,  ~~\forall t\in \{1,..., T\}  \label{eq: suc_constraint}\\
& ~~B(t)+G(t)+\lambda_e(t)\geq E(t),  ~~\forall t\in \{1,..., T\} \nonumber \\
& ~~G(t)\geq 0, ~~\forall t\in \{1,..., T\} \\
& ~~ 0\leq \rho(t)\leq 1,  ~~\forall t\in \{1,..., T\}, \label{eq: rho}\\
& ~~B(t+1)=\min\left(B(t)+ \lambda_e(t)+G(t)-E(t),C\right), ~~\forall t\in \{1,..., T-1\},\nonumber
\end{align}
where $E(t)$ is determined by $\rho(t)$ and is given in (\ref{eq: E_t}). 
However, for a general $\alpha$,  due to the complicated expression of  $\mathcal{P}_{suc}(t)$ in (\ref{eq: P_suc_general}), 
it is generally difficult to analyze the effects of $\rho(t)$ on the successful downlink transmission probability constraint.
The following proposition simplifies the successful downlink transmission probability constraint and obtains an equivalent constraint of $\rho(t)$.

\begin{proposition} \label{proposition: rho_constraint}
 When  $\alpha=4$, for each horizon $t$, there exists a  minimum required BSs' active operation probability 
 $\rho_{min}(t)=\max\left(\frac{\lambda_m(t)B}{\lambda_B W}v\frac{1-\epsilon}{\epsilon}, \frac{g_0 \sqrt{\beta \sigma^2}}{\pi^{\frac{3}{2}}\lambda_B\sqrt{P_B}}\right)$ 
 as a non-decreasing function of time-varying MT density $\lambda_m(t)$,  where  $v$ is given in Proposition \ref{proposition: P_suc} and $g_0$ is the unique solution to
$g_0 Q\left(\frac{g_0}{2 \pi}\right) =(1-\epsilon)\exp \left(-\frac{g_0^2}{4
\pi}\right)$, such that   $\mathcal{P}_{suc}(t)\geq 1-\epsilon$ is equivalent to  $\rho(t)\geq \rho_{min}(t)$.
 \end{proposition}

Proposition  \ref{proposition: rho_constraint} is proved by  using the same method  as in our previous work \cite{Che.JSAC.15}, which is omitted here for brevity.
It is noted  that when the MT density and thus the required total bandwidth $\lambda_m(t)B$ of the MTs  increase, 
due to the decreased number of bands $\delta(t)$ for frequency reuse, as given in (\ref{eq: delta}), the downlink interference level increases. 
In this case, it is observed  from Proposition  \ref{proposition: rho_constraint} that the minimum required BSs' active operation probability $\rho_{min}(t)$ also increases, 
such that more BSs need to be active to increase the desired signal strength at the MT, so as to assure a high $\mathcal{P}_{suc}(t)$. 
Due to the variations of MT density $\lambda_m(t)$, it is also observed that $\rho_{min}(t)$ is also varying over time in general. 
Moreover, similar to \cite{Che.JSAC.15},   the noise power $\sigma^2\neq 0$ in the  expression of  $\rho_{min}(t)$ provides a 
valid minimum required active operation probability for each BS,   which is important to assure  a sufficiently large $\mathcal{P}_{suc}$ in a noise-dominant network. 

Therefore, for the ease of analysis, we focus on the case of $\alpha=4$ in the sequel.\footnote{The value of $\alpha$ does not affect the main results of this paper.  
Moreover, for other cases with $\alpha\neq 4$,  it is easy to verify from (\ref{eq: P_suc_general}) that there always exist  feasible regions for $\rho(t)$ 
such that $\mathcal{P}_{suc}\geq 1-\epsilon$ is guaranteed, though it is more difficult to exactly define these feasible regions. 
Still,  our later  analysis  for studying the on-grid energy cost minimization problem applies.  } 
By applying Proposition~\ref{proposition: rho_constraint}, we find an equivalent problem to problem (P1) as follows 
\begin{align}
\textrm{(P2)}:~~~\mathop{\mathrm{min.}}_{\rho(t), G(t)}&~~   \sum_{t=1}^{T} a(t)G(t)  \\
\mathrm{s.t.}
& ~~ G(t)\geq \max(E(t)-B(t)-\lambda_e(t),0),  ~~\forall t\in \{1,..., T\}, \label{eq: G_constraint}\\
& ~~ \rho_{min}(t)\leq \rho(t)\leq 1,  ~~\forall t\in \{1,..., T\}, \label{eq: rho_constraint}\\
& ~~B(t+1)=\min\left(B(t)+ \lambda_e(t)+G(t)-E(t),C\right), ~~\forall t\in \{1,..., T-1\}, \nonumber
\end{align}
where the new constraint  (\ref{eq: G_constraint}) is obtained by combining the energy demand constraint  $B(t)+G(t)+\lambda_e(t)\geq E(t)$ 
and the non-zero on-grid energy constraint $G(t)\geq 0$ in problem (P1), and the other new  constraint (\ref{eq: rho_constraint}) is obtained 
by combining the constraints (\ref{eq: suc_constraint}) and (\ref{eq: rho}) in problem (P1), as well as Proposition \ref{proposition: rho_constraint}.
 To avoid the trivial case where the load is too high to meet even when  all BSs are activated,  we assume $\rho_{min}(t)\leq 1$ in the sequel.
 In the next section, we focus on solving problem (P2) by using the DP technique.

 \section{Policy Design for On-grid Energy Cost Minimization} \label{section: policy_design}
This section studies the optimal on-grid energy cost minimization policy for solving problem (P2).
The optimal on-grid energy minimization policy consists of two parts. One is the optimal BS on/off policy, 
which decides the optimal BSs' active operation probability $\rho(t)$ in each horizon $t$. 
The other is the optimal on-grid energy purchase policy, which decides the amount of on-grid energy $G(t)$ to purchase in each horizon $t$.
In the following, by studying the cost-to-go function that is defined based on the Bellman's equations \cite{DP}, 
we first investigate the optimal BS on/off policy, and then focus on designing the optimal on-grid energy purchase policy.

 \subsection{Cost-To-Go Function}
In this subsection, we  define  the cost-to-go function for problem (P2).
Due to the fluctuations of the on-grid energy prices, the MT density, as well as the renewable energy arrival rates over time, 
the decision of $G(t)$ and $\rho(t)$ in each horizon $t$ are  related to both current and future values of the on-grid energy prices, 
the MT density, and the renewable energy arrival rates, as well as the current storage level $B(t)$ at the CES.
We thus define the system state in horizon $t$ as 
${\bf\Lambda}(t)=\left(B(t), \left(\lambda_e(t),...,\lambda_e(T)\right), \left(a(t),...,a(T)\right), \left(\lambda_m(t),...,\lambda_m(T) \right)\right)$.
Based on the Bellman's equations \cite{DP}, for a given system state  ${\bf\Lambda}(t)$ in horizon $t$, 
we then define the cost-to-go function $J_t({\bf\Lambda}(t))$ for problem (P2)  as
\begin{equation}
   J_t({\bf\Lambda}(t)) =\left\{
   \begin{array}{l}
  \mathop{\mathrm{min}}_{\rho_{min}(T) \leq \rho(T) \leq 1,\atop G(T)\geq \max(E(T)-B(T)-\lambda_e(T),0) }  a(T) G(T),~~~~~~~~~~~~~~~~~~~~~~~~~\!\textrm{if~} t=T, \\
  \mathop{\mathrm{min}}_{\rho_{min}(t) \leq \rho(t) \leq 1,\atop G(t)\geq \max(E(t)-B(t)-\lambda_e(t),0)}   a(t)G(t)+  J_{t+1}({\bf\Lambda}(t+1)|{\bf\Lambda}(t)),~\textrm{if~} 1\leq t<T.
   \end{array}
  \right. \label{eq: J_t}
\end{equation}
  When $1\leq t<T$,   the first term $a(t) G(t)$ in  (\ref{eq: J_t}) represents the instantaneous  on-grid  energy cost   in the current horizon $t$, 
  and the second term $J_{t+1}({\bf\Lambda}(t+1)|{\bf\Lambda}(t))$ gives the minimized  on-grid energy cost over all the future horizons from $t+1$ to $T$, 
  where  for a given ${\bf\Lambda}(t)$ and thus $B(t)$, $B(t+1)$ in ${\bf\Lambda}(t+1)$ is updated according to (\ref{eq: B_t}). 
  When $t=T$, only the instantaneous  on-grid  energy cost  is considered in  (\ref{eq: J_t}).
By finding the cost-to-go function $J_t({\bf\Lambda}(t))$ in each horizon $t$, one can equivalently obtain the optimal BS on/off policy and the optimal on-grid energy purchase policy, 
which gives the optimal $\rho^*(t)$ and $G^*(t)$ in each horizon $t$ for problem (P2), respectively.  
Moreover, unlike \cite{Fu.Infocomm.03},  due to the update of the storage level  in (\ref{eq: B_t}) that correlates the optimal $G^*(t)$ and $\rho^*(t)$ over time,  
a joint design of the optimal BS on/off policy and the optimal on-grid energy purchase policy is generally required.

In the next two subsections, we first study the optimal BS on/off policy, under a given  on-grid energy purchase policy. 
Then by substituting the obtained optimal BS on/off policy into the cost-to-go function (\ref{eq: J_t}), we study the optimal on-grid energy purchase policy.

\subsection{Optimal BS On/Off Policy}
This subsection studies the optimal BS on/off policy  for problem (P2), by supposing that an arbitrary  on-grid energy purchase policy is given, 
i.e., the microgrid knows the amount of on-grid energy $G(t)$ to purchase for a given system state ${\bf\Lambda}(t)$ in each horizon $t$.

We first study the impact of storage level $B(t)$ on the cost-to-go function $J_t({\bf \Lambda}(t))$ in the following   lemma.
\begin{lemma}\label{lemma: J_B}
 The cost-to-go function $J_t({\bf \Lambda}(t))$ in horizon $t$ is non-increasing over the current storage level $B(t)$.
\end{lemma}
\begin{IEEEproof}
 Lemma \ref{lemma: J_B} is proved based on the mathematical induction method. First, at the last horizon $t=T$, 
 the optimal $G^*(T)$ is given as $G^*(T)=\max(E(T)-B(T)-\lambda_e(T),0)$. As a result,  $G^*(T)$ and thus $J_T({\bf \Lambda}(T))=a(T)G^*(T)$ 
 is non-increasing over $B(T)$. Next, suppose at horizon $t+1$, $J_{t+1}({\bf \Lambda}(t+1))$  is non-increasing over $B(t+1)$. 
 Then, at horizon $t$, when $B(t)$ increases, since $B(t+1)$ is non-decreasing from (\ref{eq: B_t}), the future cost $J_{t+1}({\bf \Lambda}(t+1)|{\bf \Lambda}(t))$ is non-increasing. 
 Moreover, similar to the case in horizon $T$, the current cost $a(t)G(t)$ is also non-increasing over $B(t)$. 
 Therefore, by combing the current and future cost, we find that  $J_{t}({\bf \Lambda}(t))$  is also non-increasing over $B(t)$. Lemma \ref{lemma: J_B} thus follows.
\end{IEEEproof}

Next, by studying the impact of $\rho(t)$ in the current horizon on the storage level $B(t+1)$ in the next horizon, we obtain the following lemma.
\begin{lemma}\label{lemma: J_rho}
 For any given on-grid energy purchase policy, the cost-to-go function $J_t({\bf \Lambda}(t))$ in horizon $t$ is non-decreasing over $\rho(t)$.
\end{lemma}

The proof of Lemma \ref{lemma: J_rho} can be easily obtained by applying Lemma \ref{lemma: J_B} and  the fact that for any given storage level $B(t)$ and purchase amount $G(t)$
in the current horizon, the storage level $B(t+1)$ in the next horizon is non-increasing over  $\rho(t)$ in the current horizon  according to (\ref{eq: E_t}) and (\ref{eq: B_t}), and thus is omitted here for brevity.

Finally, as a straightforward result  from Lemma \ref{lemma: J_rho}, to minimize the overall on-grid energy cost in all $T$ horizons, we obtain the optimal BS on/off policy as follows.
\begin{proposition}[Optimal BS on/off policy]  \label{proposition: optimal_BS_on_off}
The optimal $\rho^*(t)$ in each horizon $t$ for problem (P2) is given by
\begin{equation}
 \rho^*(t)=\rho_{min}(t),  ~~\forall t\in\{1,...,T\},\label{eq: optimal_rho}
\end{equation}
where $\rho_{min}(t)$ is given in Proposition \ref{proposition: rho_constraint} and depends on time-varying MT density $\lambda_m(t)$.
\end{proposition}
\begin{remark}
From Proposition \ref{proposition: optimal_BS_on_off}, since a small $\rho(t)$ yields a low energy demand $E(t)$ of the BSs, and thus a low on-grid energy cost in general,
the optimal $ \rho^*(t)$ in each horizon $t$ is always selected as the minimum required $\rho_{min}(t)$ which is just sufficient to assure  the required downlink transmission quality.
Based on $ \rho^*(t)$, each BS then independently decides its on/off status  in each horizon $t$. It is worth noting that 
since the traffic loads of the inactive BSs are handed over to their neighboring active BSs, as described in Section II, 
although the exact on/off status of the BSs are determined independently, the network-level downlink transmission quality 
can still be assured with  $ \rho^*(t)=\rho_{min}(t)$. 
It is also  noted  from Proposition \ref{proposition: rho_constraint} that the optimal $ \rho^*(t)=\rho_{min}(t)$
generally varies over time  to adapt to the variations of the MT density $\lambda_m(t)$.

\end{remark}

 \subsection{Optimal On-Grid Energy Purchase Policy}
 This subsection studies the on-grid energy purchase policy for problem (P2). By substituting the optimal $\rho^*(t)$ into  (\ref{eq: J_t}),
 the cost-to-go function  for problem (P2) is simplified as
 \begin{equation}
   J_t({\bf\Lambda}(t)) =\left\{
   \begin{array}{l}
  \mathop{\mathrm{min}}_{ G(T)\geq \max(E_{min}(T)-B(T)-\lambda_e(T),0) }  a(T) G(T),~~~~~~~~~~~~~~~~~~~~~~~~\!\textrm{if~} t=T, \\
  \mathop{\mathrm{min}}_{ G(t)\geq \max(E_{min}(t)-B(t)-\lambda_e(t),0)}   a(t)G(t)+  J_{t+1}({\bf\Lambda}(t+1)|{\bf\Lambda}(t)),~\textrm{if~} 1\leq t<T.
   \end{array}
  \right. \label{eq: J_t_G}
\end{equation}
  where   by substituting $\rho^*(t)=\rho_{min}(t)$ into (\ref{eq: E_t}), we obtain the minimum value of $E(t)$, denoted by $E_{min}(t)=E(t)|_{\rho(t)=\rho_{min}(t)}$, as
  \begin{equation}
   E_{min}(t)=\lambda_B \rho_{min}(t) P_{gap}+\lambda_B P_s+\frac{P_B\lambda_m(t)}{\mu}, \label{E_min_t}
  \end{equation}
and accordingly, the storage level $B(t+1)$ in system state ${\bf\Lambda}(t+1)$ is updated from $B(t)$ in system state ${\bf\Lambda}(t)$ as
\begin{equation}
 B(t+1)=\min\left(B(t)+ \lambda_e(t)+G(t)-E_{min}(t),C\right).  \label{eq: B_t_G}
\end{equation}
 However, unlike the design of optimal BS on/off policy, due to the fluctuations of the on-grid energy price $a(t)$ over time, 
 the cost-to-go function $J_t({\bf\Lambda}(t))$ in (\ref{eq: J_t_G})
 is generally neither non-increasing nor non-decreasing over $G(t)$ in each horizon $t$. Although based on (\ref{eq: J_t_G}), 
 one can always find the optimal $G^*(t)$, the complexity  of finding $G^*(t)$ in each horizon increases exponentially over the total number of horizons $T$, 
 due to the curse of dimensionality for the DP problem.
 As a result, in the following, to obtain tractable and insightful on-grid energy purchase policy for problem (P2), 
 we focus on designing a suboptimal on-grid energy purchase policy, by studying the optimal  solutions of $G^*(t)$ for the last horizon $T$ and the horizon $T-1$. 
 We also assume the storage capacity $C$ is sufficiently large with $C\geq \max\left(E_{min}(1), ..., E_{min}(T)\right)$, 
 such that the BSs' demanded energy for each horizon can be properly stored.  

 First, we look at the last horizon with $t=T$, and assume the  system state ${\bf\Lambda}(T)=\left(B(T), \lambda_e(T),a(T),\lambda_m(T)\right)$ in horizon $T$ is given. 
 From  (\ref{eq: J_t_G}), it is easy to find that the optimal $G^*(T)$ in the last horizon $T$ is given by
 \begin{equation}
  G^*(T)=\max(E_{min}(T)-B(T)-\lambda_e(T),0).  \label{eq: optimal_G_T}
 \end{equation}
It is also noted that the optimal $G^*(T)$ is a myopic solution, which minimizes the instantaneous on-grid energy cost in the current horizon. 
Moreover, when $G^*(T)$ is myopic, it is noted that the optimal $\rho^*(T)$ that minimizes $E(T)$ and thus the instantaneous on-grid energy cost $a(T)G(T)$ is also a myopic solution.

 Next, we consider the horizon with $t=T-1$, and  assume the system state in horizon $T-1$, i.e., 
 ${\bf\Lambda}(T-1)=\left(B(T-1), \left(\lambda_e(T-1),\lambda_e(T)\right),\left(a(T-1),a(T)\right),\left(\lambda_m(T-1),\lambda_m(T)\right)\right)$, is given. 
 We also assume the harvested renewable energy in horizon $T-1$ is reasonably small with $\lambda_e(T-1)<C+E_{min}(T-1)-B(T-1)$, as compared to the storage capacity and the 
 energy demand of all the BSs.
 Based on (\ref{eq: J_t_G}), we jointly consider the impact of $G(T-1)$ on the instantaneous on-grid energy cost in the current horizon $T-1$ 
 and the future on-grid energy cost  $J_{T}({\bf\Lambda}(T)|{\bf\Lambda}(T-1))$, 
 where the storage level $B(T)$ in the next horizon $T$ is updated from $B(T-1)$ according to (\ref{eq: B_t_G}), and obtain the optimal $G^*(T-1)$ in the following proposition.
 \begin{proposition} \label{proposition: optimal_G_T-1}
  Given the system state ${\bf\Lambda}(T-1)$ in horizon $T-1$,   the optimal $G^*(T-1)$ for problem (P2) is given as follows:
  \begin{itemize}
   \item \emph{Large current storage regime with $B(T-1)\geq E_{min}(T)+E_{min}(T-1)-\lambda_e(T-1)-\lambda_e(T)$}: we have
   \begin{equation}
    G^*(T-1)=\max(E_{min}(T-1)-B(T-1)-\lambda_e(T-1),0).  \label{eq: optimal_G_T-1_myopic}
   \end{equation}

   \item \emph{Low current storage regime with  $B(T-1)<E_{min}(T)+E_{min}(T-1)-\lambda_e(T-1)-\lambda_e(T)$}:
          \begin{itemize}
           \item \emph{High future renewable energy regime with  $\lambda_e(T)\geq E_{min}(T)$}: $G^*(T)$ is  also given by (\ref{eq: optimal_G_T-1_myopic}).
           \item \emph{Low future renewable energy regime with  $\lambda_e(T)< E_{min}(T)$}: we have
           \begin{equation}
               G^*(T-1)=\left\{
            \begin{array}{l}
              \max(E_{min}(T-1)-B(T-1)-\lambda_e(T-1),0),~~~~~~~~~~~~~~~\textrm{if~} a(T-1)\geq a(T), \\
              E_{min}(T)+E_{min}(T-1)-\lambda_e(T-1)-\lambda_e(T)-B(T-1),~\textrm{if~} a(T-1)< a(T).
            \end{array}
              \right.
           \end{equation}

          \end{itemize}

  \end{itemize}

 \end{proposition}
\begin{IEEEproof}
 Please refer to Appendix A.
\end{IEEEproof}

\begin{observation}
 From  Proposition  \ref{proposition: optimal_G_T-1},  the decision of $G^*(T-1)$ is jointly determined by the fluctuations of MT density, on-grid energy prices, the renewable energy
 arrival rates in both current and future horizons.  
 Moreover, the on-grid energy  $G^*(T-1)$ is  over-purchased only   when all the following three conditions are satisfied:
 \begin{itemize}
  \item[1)] The current storage level $B(T-1)$ is within the low current storage regime;
  \item[2)] The future renewable energy $\lambda_e(T)$   alone  cannot  to satisfy the demanded energy $E_{min}(T)$ in the future;
  \item[3)] The current on-grid energy price is lower than the future with $a(T-1)< a(T)$.
 \end{itemize}
 When all the three conditions are satisfied, the microgrid purchases more energy than the demanded in the current horizon, 
 so as to ensure both current and future energy demands can be satisfied.
 In addition, it is also observed that  if the on-grid energy prices $a(T-1)=a(T)$,  $G^*(T-1)$ is not affected by the fluctuations of the on-grid energy prices 
 and becomes a myopic solution in all cases of Proposition  \ref{proposition: optimal_G_T-1},  just as $G^*(T)$  in (\ref{eq: optimal_G_T}).

\end{observation}

Finally, we extend  the above observations for  the optimal  $G^*(t)$ obtained in horizon $T$ and horizon $T-1$ to the on-grid energy purchase decision 
in an arbitrary horizon $t\in\{1,...,T\}$, and propose a suboptimal on-grid energy purchase policy for problem (P2) in the following, 
by jointly considering the impact of storage capacity $C$.

\begin{proposition}[Suboptimal on-grid energy purchase policy] \label{proposition: suboptimal_G_policy}
Denote $G^{\star}(t)$  as the suboptimal solution of the purchased on-grid energy to problem (P2) in each horizon $t\in\{1,...,T\}$.
The microgrid  over-purchases the on-grid energy   in horizon $t$ if all of the three following conditions are satisfied at the same time:
\begin{enumerate}
 \item \emph{Low current storage level:} $\min \left(B(t)-E_{min}(t)+\lambda_e(t),C \right)<\sum_{k=t+1}^T E_{min}(k)- \sum_{k=t+1}^T \lambda_e(k)$;
 \item \emph{Low future renewable energy:} $\sum_{k=t+1}^T E_{min}(k)- \sum_{k=t+1}^T \lambda_e(k)\geq 0$;
 \item \emph{Low current on-grid energy price:} $a(t)< \min\left(a(t+1),...,a(T) \right)$.
\end{enumerate}

Let $\omega(t)= \min\left(\sum_{k=t+1}^T E_{min}(k)- \sum_{k=t+1}^T \lambda_e(k),C\right)$. Given the system state ${\bf \Lambda(t)}$ in horizon $t$, 
$G^{\star}(t)$ is given by
\begin{equation}
   G^{\star}(t) \!=\!\left\{
   \begin{array}{l}
 \! \!\max \left(\omega(t)+E_{min}(t)-\lambda_e(t)-B(t), 0\right),~ \textrm{if~conditions~1)-3)~hold}, \\
 \!\! \max(E_{min}(t)-B(t)-\lambda_e(t),0),~~~~~~~~~~    \textrm{otherwise}.
   \end{array}
  \right. \label{eq: G_suboptimal_t}
\end{equation}

\end{proposition}

From Proposition \ref{proposition: suboptimal_G_policy},  the microgrid over-purchases the low-price energy in the current horizon  for future usage 
only when the current storage and future renewable supply are both low. It is also easy to verify that $G^{\star}(t)=G^{*}(t)$ when $t=T$ or $t=T-1$. 
Moreover, when $a(t)=a(t')$, $\forall t, t' \in \{1,...,T\}$ and $t\neq t'$,  $G^{\star}(t)$ becomes the myopic solution $\max(E_{min}(t)-B(t)-\lambda_e(t),0)$ for all horizons, 
which is similar to the case with $t=T-1$ in Proposition \ref{proposition: optimal_G_T-1}.   
Moreover, it is also easy to find that  as compared to the DP-based approach to search the optimal $G^{*}(t)$ in each horizon by applying (\ref{eq: J_t_G}),
the proposed suboptimal on-grid energy purchase policy is of low complexity to implement. 
We will validate the performance of the proposed suboptimal on-grid energy purchase  policy for multi-horizon  in Section~VI.

 To conclude, by combining the optimal suboptimal BS on/off policy and the suboptimal on-grid energy purchase  policy, 
 given in Proposition \ref{proposition: optimal_BS_on_off} and Proposition  \ref{proposition: suboptimal_G_policy}, respectively,  
 the on-grid energy cost minimization problem  (P2) is solved.

\section{Simulation Results}

This section studies the proposed network-level on-grid energy cost minimization policy by providing extensive numerical and simulation results.
 We first   illustrate the adaptation of $\rho(t)$ and $G(t)$ over horizons  by applying the proposed  policy. 
 We then validate the performance of the proposed suboptimal on-grid energy purchase policy. After that, we study the effects of the prediction errors 
 in terms of the network profiles  $\lambda_e(t)$'s, $\lambda_m(t)$'s, and $a(t)$'s. At last, we conduct a large-scale network-level simulation 
 to compare the performance of the proposed policy with two reference schemes.     

 \begin{figure}[!t!b!h]
\setlength{\abovecaptionskip}{-0.15in}
\centering
\DeclareGraphicsExtensions{.eps,.mps,.pdf,.jpg,.png}
\DeclareGraphicsRule{*}{eps}{*}{}
\includegraphics[angle=0, width=0.65\textwidth]{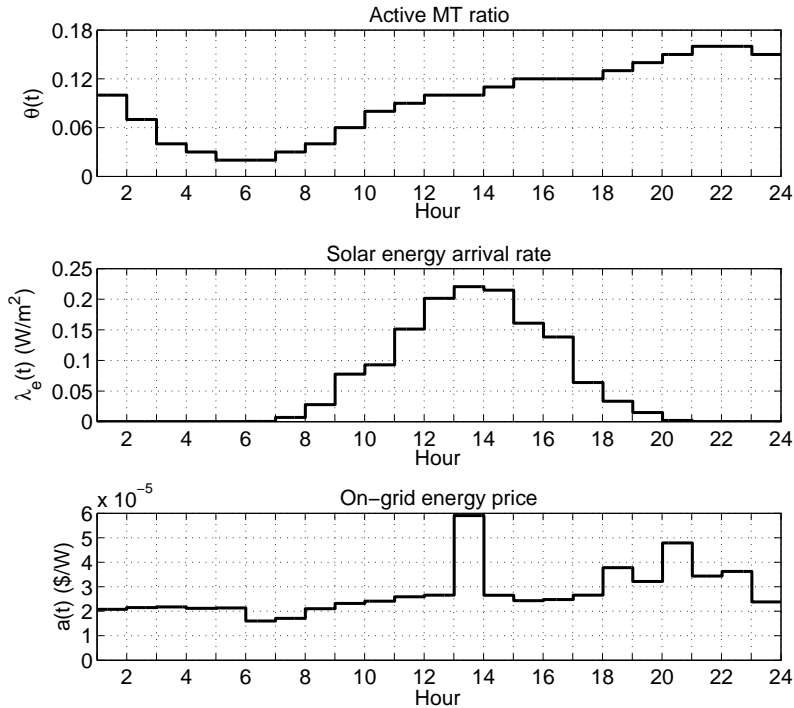}
\caption{Network profiles on the variations of $\lambda_m(t)$, $\lambda_e(t)$, and $a(t)$  for a single day.}
\label{fig: network_parameters}
\vspace{-0.1in}
\end{figure}
 Without specified otherwise, we consider the following simulation parameters in this section. 
 We set the basic energy consumptions of each BS in the active and inactive modes as $P_a=130$W and  $P_s=75$W, respectively,  
 the transmit power of each active BS   as $P_B=20W$, and the power amplifier parameter as $\mu=21.3\%$ \cite{Earth}.
 We also set the noise power $\sigma^2=10^{-9}$W,  the targeted SINR level  $\beta=2$, the path-loss exponent $\alpha=4$, 
 and the maximum outage $\epsilon=0.05$ to ensure a sufficiently high downlink successful transmission probability $\mathcal{P}_{suc}(t)$ in each horizon. 
 According to  the real data measured  in \cite{Earth}, \cite{E_data},  and \cite{price},
 we consider the variations of  the MT density $\lambda_m(t)$, the solar energy arrival rate $\lambda_e(t)$, as well as the on-grid energy price $a(t)$ over a single day, 
 and show them in Fig.~\ref{fig: network_parameters}. 
 In particular,  instead of $\lambda_m(t)$, Fig.~\ref{fig: network_parameters} shows the active MT ratio $\theta(t)$ for each horizon $t$. 
 By denoting the total MT density in the network as $\lambda_m^{all}>0$, we can obtain the MT density in each horizon as $\lambda_m(t)=\lambda_m^{all}\theta(t)$. 
 We set $\lambda_m^{all}=8\times 10^{-3}/\textrm{m}^2$, and the BS density as $\lambda_B=5\times 10^{-4}/\textrm{m}^2$.  
 We also set the storage capacity $C$ at the CES, given in (\ref{eq: B_t_G}), as $C=0.2$ (W/$\textrm{m}^2$).

\subsection{Illustration for Dynamic Network Operation}
In this subsection,  we illustrate the  dynamic adaptation of the BSs' active operation probability  and the amount of on-grid energy to purchase over  horizons, 
by applying the optimal BS on/off policy and the suboptimal on-grid energy purchase policy, given in Proposition \ref{proposition: optimal_BS_on_off} and Proposition \ref{proposition: suboptimal_G_policy}, respectively.

  \begin{figure}
\setlength{\abovecaptionskip}{-0.15in}
\centering
\DeclareGraphicsExtensions{.eps,.mps,.pdf,.jpg,.png}
\DeclareGraphicsRule{*}{eps}{*}{}
\includegraphics[angle=0, width=0.8\textwidth]{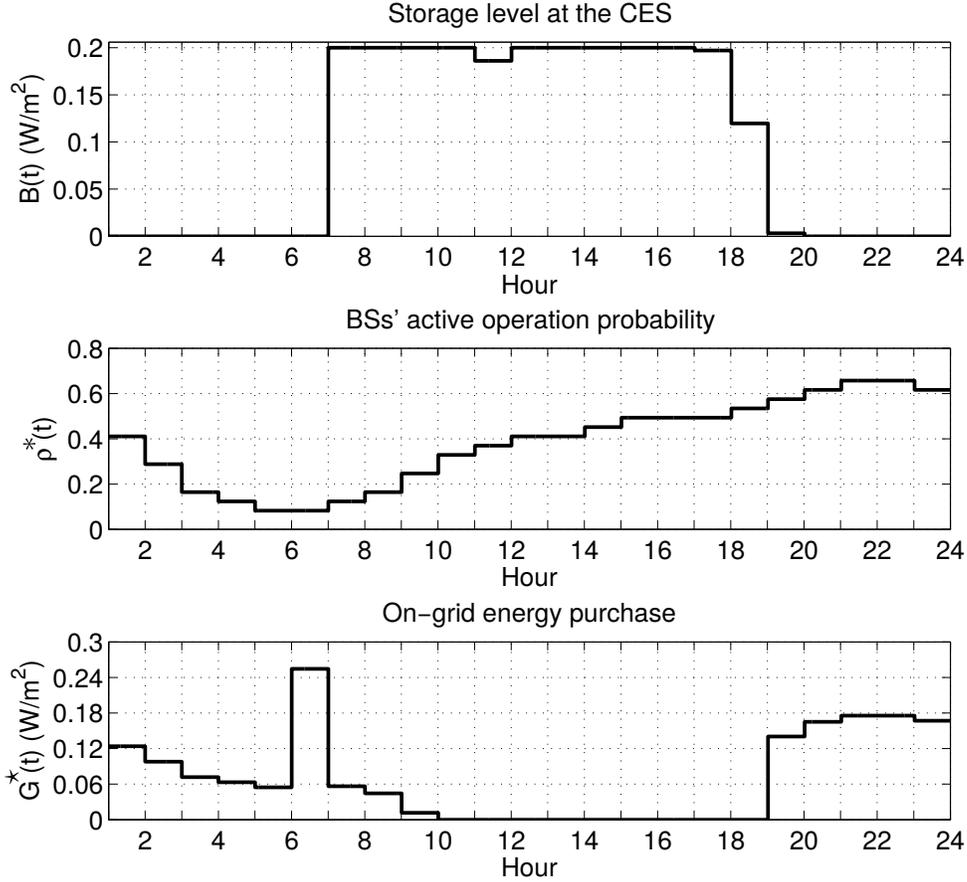}
\caption{Adaptations of $\rho^*(t)$ and $G^{\star}(t)$ and the variation of $B(t)$ for a single day.}
\label{fig: rho}
\vspace{-0.1in}
\end{figure}

By applying the network profiles given in Fig.~\ref{fig: network_parameters},    Fig.~\ref{fig: rho} shows the  variations of the  storage level $B(t)$, 
as well as the adaptations of the  BSs' active operation probability  and the amount of purchased on-grid energy  to these network profiles  over $T=24$ hours.
From Proposition \ref{proposition: optimal_BS_on_off}, the optimal $\rho^*(t)=\rho_{min}(t)$ in each horizon, 
where $\rho_{min}(t)$ given in Proposition \ref{proposition: rho_constraint} varies monotonically over the MT density $\lambda_e(t)$. 
It is thus observed from Fig.~\ref{fig: rho} that $\rho^*(t)$ follows the exact variation trend of  $\lambda_e(t)$ in Fig.~\ref{fig: network_parameters}.
It is also observed that when the solar energy arrival rate is substantially large from $t=7$ to $t=19$ in Fig.~\ref{fig: network_parameters}, 
the storage level $B(t)$ in  Fig.~\ref{fig: rho}  is also quite high, and achieves the capacity of the CES in most horizons  from $t=7$ to $t=19$. 
However, when the   solar energy arrival rate is approaching to zero from $t=1$ to $t=6$ and from $t=20$ to $t=24$ in Fig.~\ref{fig: network_parameters}, 
the storage level $B(t)$ in  Fig.~\ref{fig: rho} is also almost zero in each of these horizons.
Moreover,  from $t=1$ to $t=5$, since the storage level  is almost zero in Fig.~\ref{fig: rho}, 
but the MT density is not low and the on-grid energy price  is non-decreasing in Fig.~\ref{fig: network_parameters}, 
it is observed from Fig.~\ref{fig: rho} that the purchased on-grid energy follows the variation trends of the MT density and thus is myopic solution in each of these horizons. 
At horizon $t=6$, it is observed from Fig.~\ref{fig: network_parameters} that $a(6)$ at $t=6$ is the lowest value among all the on-grid energy prices over $T$ horizons. 
Thus, from the suboptimal on-grid energy purchase policy given in Proposition \ref{proposition: suboptimal_G_policy}, 
it is noted that the purchased on-grid energy $G(6)$ at $t=6$ achieves the highest value among all the $T$ horizons. 
From $t=7$ to $t=19$, due to the large storage level $B(t)$, the purchased on-grid energy is decreasing in these horizons and becomes zero 
from $t=10$ to $t=18$ in  Fig.~\ref{fig: rho}. At last, from $t=19$ to $t=24$, due to the 
almost zero storage level in  Fig.~\ref{fig: rho} as well as the large MT density in Fig.~\ref{fig: network_parameters}, 
the purchased   on-grid energy is largely increased in these horizons  in  Fig.~\ref{fig: rho}.

\subsection{Suboptimal Policy Performance Validation}  \label{section: sub_validate}
 \begin{table}[!h!b!t]
\centering
\setlength{\belowcaptionskip}{-0.1in}
\caption{Comparison of  total on-grid energy cost $ \sum_{t=1}^{T} a(t)G(t)$ under optimal and suboptimal polices.} 
\begin{tabular}{|c|c|c|c|c|c|}
\hline 
 Number of horizons & $T=1$ & $T=2$ & $T=3$ &$T=4$ & $T=5$ \\
 \hline
Optimal ($\times 10^{-6}$\$/$\textrm{m}^2$)& 2.1004935 & 3.6441487 &5.005297 & 6.1624714 &7.0429205 \\
 \hline
 Suboptimal  ($\times 10^{-6}$\$/$\textrm{m}^2$)& 2.1004935 & 3.6441487 &5.005329 & 6.1624757 &7.0429250 \\
   \hline
 Absolute error  ($\times 10^{-6}$\$/$\textrm{m}^2$)& 0& 0& $3.2\times 10^{-5}$& $4.3\times 10^{-5}$& $4.5\times 10^{-5}$\\
   \hline
\end{tabular}
\vspace{-0.1in}
\end{table}
Since the proposed BS on/off policy in Proposition \ref{proposition: optimal_BS_on_off} is optimal, 
the  suboptimality of the proposed   on-grid energy minimization policy is caused by the suboptimality 
of the proposed on-grid energy purchase policy in Proposition \ref{proposition: suboptimal_G_policy}. 
Thus, by applying the optimal BS on/off policy, we  compare the performance of the suboptimal    
on-grid energy purchase policy in Proposition \ref{proposition: suboptimal_G_policy} with the optimal on-grid energy purchase policy that is obtained by exhaustive search.  
To search the optimal $G^*(t)$, we first find the search range of $G(t)$ in each horizon $t$. 
From (\ref{eq: J_t_G}),  the minimum required $G(t)$ is given by $ G_{min}(t)=\max(E_{min}(t)-B(t)-\lambda_e(t),0)$. 
It is also easy to obtain that to save the on-grid energy cost, the maximum required $G(t)$ is 
given by $G_{max}(t)=\sum_{k=t}^{T}E_{min}(k)-\sum_{k=t}^{T}\lambda_e(k)$, 
which can   just satisfy the energy demand of all the BSs from the current to all future horizons. 
Then, we apply the DP technique from \cite{DP} to search the optimal $G^*(t)\in[G_{min}(t),G_{max}(t)]$ in each horizon $t$, 
where  the range of $[G_{min}(t),G_{max}(t)]$ is quantized with a sufficiently small step-size  $1.0\times 10^{-8}$.
  In Table I, we compare $ \sum_{t=1}^{T} a(t)G^*(t)$ that is obtained by the optimal search 
  with   $ \sum_{t=1}^{T} a(t)G^{\star}(t)$ that is obtained by applying the proposed suboptimal  
  on-grid energy purchase policy in  Proposition \ref{proposition: suboptimal_G_policy}.
Due to the curse of dimensionality to apply the DP technique for searching the optimal $G^*(t)$, 
we only consider the total number of horizons $T\in \{1,..,5\}$ in Table I, using network profiles   
starting from $t=2$  in Fig.~\ref{fig: network_parameters} as an example.
It is observed from Table I that both the optimal and suboptimal total on-grid energy cost increases over $T$. 
It is also observed that  as   the number of horizons $T$ increases,  since the  number of    optimal  $G^*(t)$  and  
suboptimal $G^{\star}(t)$ with different values  in each horizon $t$  also increases in general,  
  the absolute error, given by $\sum_{t=1}^{T} a(t)G^{\star}(t)-\sum_{t=1}^{T} a(t)G^*(t)$, 
  increases over $T$ accordingly. However, it is observed that the absolute error at each $T$ is quite small and is within $4.5\times10^{-11}$. 
As a result, the proposed suboptimal
on-grid energy purchase policy in Proposition \ref{proposition: suboptimal_G_policy} and thus 
the proposed suboptimal on-grid energy minimization policy in Section \ref{section: policy_design}
can achieve near-optimal performance.

 \subsection{Effects of Prediction Errors} \label{section: effects_prediction_error}
In this subsection, by applying our proposed on-grid energy cost minimization policy in Section  \ref{section: policy_design}, 
we study the effects of the prediction errors for $\lambda_e(t)$'s, $\lambda_m(t)$'s, and $a(t)$'s. 
In particular, we focus on the effects of the prediction errors for the solar energy arrival rate $\lambda_e(t)$'s, and assume $\lambda_m(t)$'s  
and $a(t)$'s in all horizons are still perfectly known. We also observe similar simulation results for the cases where prediction errors exist 
in $\lambda_m(t)$  and/or $a(t)$, which are thus omitted here for brevity.

 Specifically, we assume a prediction error $\Delta(t)$ exists in each horizon $t$. We also assume $\Delta(t)$'s are identical and independently 
 distributed zero-mean Gaussian variables over different horizons, where the standard deviation  is denoted by $\eta$. 
 Since   the prediction error $\Delta(t)$ can be large due to the  large $T=24$ under consideration, 
 based on the value of $\lambda_e(t)\in[0.01~\textrm{W}/\textrm{m}^2,0.22~\textrm{W}/\textrm{m}^2]$ shown in Fig.~\ref{fig: network_parameters}, 
 we reasonably consider a large  standard deviation for the prediction errors with $\eta=0.1$ in each horizon. 
 As a result, under the consideration of prediction errors, the microgrid adopts $\lambda_e(t)+\Delta(t)$ as the  solar energy arrival rate in each horizon $t$ 
 to decide $\rho^*(t)$ and $G^{\star}(t)$.  However, since the storage level $B(t)$ in each horizon  can be perfectly known by the microgrid,  
 the storage level $B(t+1)$ in horizon $t+1$ is still updated according to the actual solar energy arrival rate $\lambda_e(t)$ in horizon $t$.

\begin{figure}
\setlength{\abovecaptionskip}{-0.25in}
\centering
\DeclareGraphicsExtensions{.eps,.mps,.pdf,.jpg,.png}
\DeclareGraphicsRule{*}{eps}{*}{}
\includegraphics[angle=0, width=0.7\textwidth]{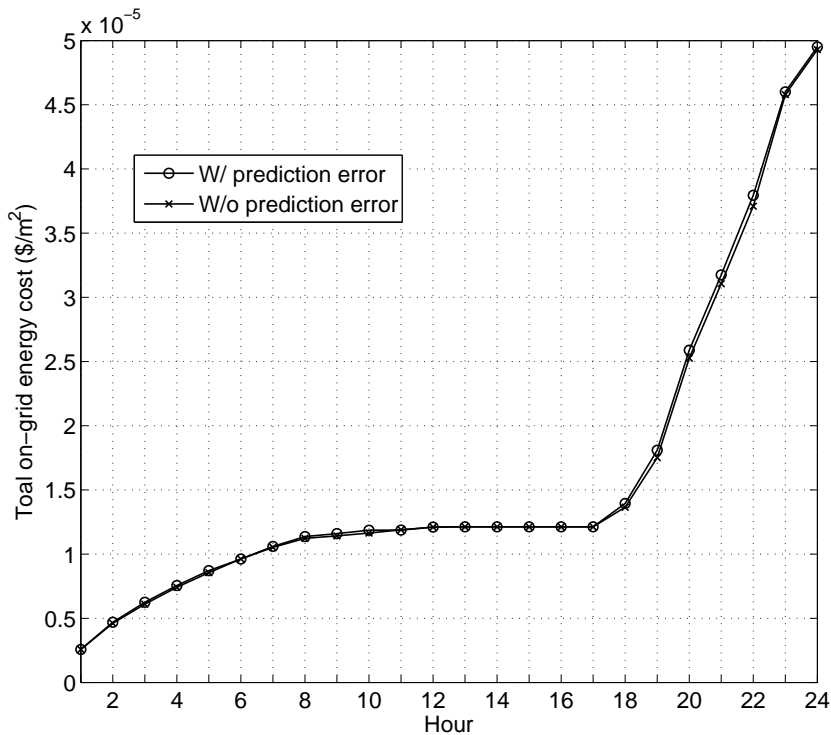}
\caption{ Effects of the predictions errors on the network performance. }
\label{fig: prediction_errors}
\vspace{-0.1in}
\end{figure}
 Fig.~\ref{fig: prediction_errors} compares the total on-grid energy cost $\sum_{t=1}^{T}a(t)G(t)$ over the number of horizons $T$ 
 under both cases with and without the prediction errors.
 For the case with prediction errors, at each $T$ in Fig.~\ref{fig: prediction_errors}, we consider  1000 realizations of the $T$ horizons with predictions errors, 
 and take the average   on-grid energy cost in $T$ horizons over the 1000 realizations as the total on-grid energy cost for comparison.
 It is observed from Fig.~\ref{fig: prediction_errors} that since we focus on the  average energy consumption, as discussed in \ref{section: energy_supplies}, 
 the network performance under the case with the prediction errors is always very close to that under the case without errors at each $T$. 
 Hence, our proposed policy is robust to the prediction errors.
It is also observed from Fig.~\ref{fig: prediction_errors} that the total on-grid energy cost generally increases with the number of horizons $T$, as expected.

\subsection{Geo-DP based Scheme versus Reference Schemes}
To show the performance of our proposed scheme based on the  Geo-DP approach, where the BSs are coordinated from a network-level to
serve the MTs,  this subsection conducts Monte Carlo simulations to compare the proposed scheme with two reference schemes,
 which are specified as follows.

 First, we consider a simple reference scheme without  BSs' mutual coordination.
 In this scheme,  each BS individually decides whether to be active or not in each horizon: 
 if there exist  MTs locating in a BS's coverage, it keeps active, or becomes inactive, otherwise. 
 We assume the BS density $\lambda_B$ is sufficiently large, such that when all the BSs are active, the downlink transmission quality is assured in the network. 
 Thus, since only BSs with no associated MTs turn to be inactive, the downlink transmission quality can be assured in this scheme. 
 The  total energy demand of the BSs  is calculated according to (\ref{eq: P_on}) and (\ref{eq: P_off}), depending on whether a BS is active or not, respectively.

  Second, we consider a reference scheme with cluster-based BS coordination, which has been proposed
  in the existing literature   to coordinate the BSs' on/off operations in a  large-scale network (see, e.g., \cite{Zhou.ICTC.13} and \cite{Niu.TCOM.14}).
  However, it is noted that due to the considered different performance metrics,  the existing  cluster-based BS coordination policy
    cannot be applied directly. For example, in \cite{Zhou.ICTC.13} and \cite{Niu.TCOM.14}, the authors considered a   blocking probability constraint 
    for the new MT's access as the QoS constraint, while we use    the SINR-based downlink transmission quality as the QoS constraint. 
    As a result, we consider the following cluster-based BS coordination scheme for comparison. Specifically,
    we assume  each cluster includes at most two BSs, and any two BSs that are located within $L>0$ meters form a cluster. 
    Each BS can belong to at most one cluster.
    If a BS does not belong to any cluster, it decides its on/off status  as that in the scheme with no BS coordination. 
    For any two BSs that belong to
    one cluster, the BS that has less associated MTs becomes inactive, while the other becomes active and also helps to serve the MTs of the inactive BS.
    The total energy consumption of all the BSs is also calculated as the same as that in the scheme with no BS coordination.   
    It is easy to verify that the downlink transmission quality 
    can be assured with a small $L$, where the MTs are always connected to the active BSs that are located closely for good communication quality.
  
  To obtain the total on-grid energy cost under our proposed scheme and the two reference schemes, we conduct a large-scale network simulation. 
   We generate $\Phi(\lambda_{B})$ for BSs and $\Phi(\lambda_m(t))$ for MTs  in  a square of
$[0\textrm{m},1000\textrm{m}]\!\times\![0\textrm{m},1000\textrm{m}]$, according to the method described in \cite{Stoyan.SG.95}.
Due to the computation complexity that increases substantially over both network size as well as the number of horizons $T$, we set $T=3$ 
and the number of slots within each horizon as  $N=1$ in this simulation, and take network profiles of $\lambda_e(t)$ and $\theta(t)$ 
from Fig.~\ref{fig: network_parameters} for  $t\in \{7,8,9\}$. For the cluster-based scheme, we use a  small $L$ with $L=100$m  to assure the 
      downlink transmission quality.  
We consider in total 20000 network realizations, and take the average value over the 20000 realizations as the network performance for each considered scheme.
For simplicity, we assume identical  on-grid energy price  with $a(t)=1$ (\$/W) in each horizon $t$. 
In this case,  it is easy to find that the on-grid energy purchase decision  for all three considered schemes is always myopic in each horizon from Section \ref{section: policy_design}. 
Similar performance can also be observed for all three schemes with non-identical on-grid energy prices, which are thus omitted here for brevity.

\begin{figure}
\setlength{\abovecaptionskip}{-0.15in}
\centering
\DeclareGraphicsExtensions{.eps,.mps,.pdf,.jpg,.png}
\DeclareGraphicsRule{*}{eps}{*}{}
\includegraphics[angle=0, width=0.7\textwidth]{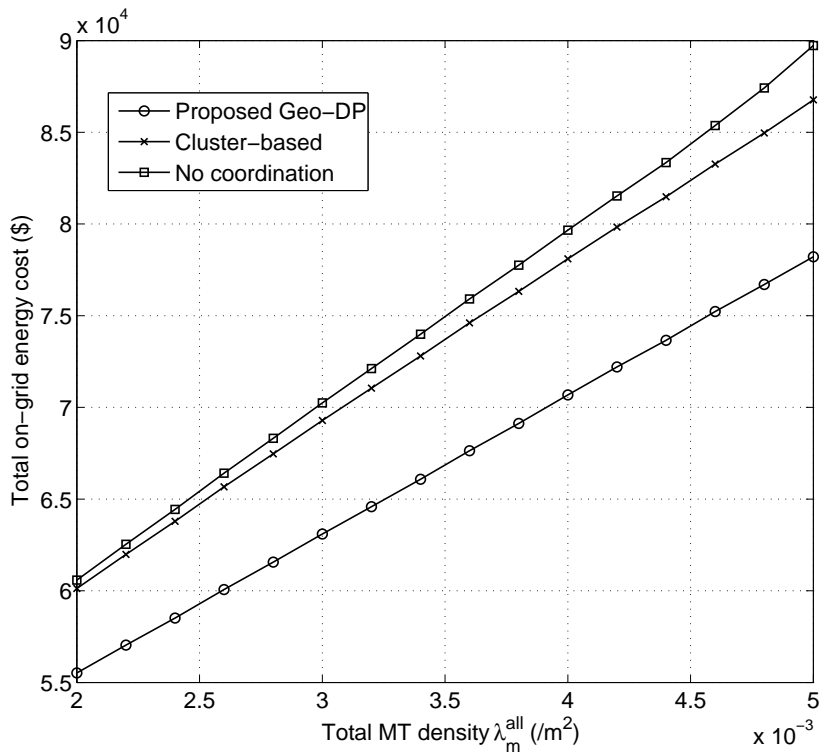}
\caption{Network performance comparison with two reference schemes.  }
\label{fig: scheme_comparison}
\vspace{-0.1in}
\end{figure}

Fig.~\ref{fig: scheme_comparison} compares the total on-grid energy cost across $T$ horizons over the total MT density $\lambda_m^{all}$.
It is observed  from Fig.~\ref{fig: scheme_comparison} that as the total MT density $\lambda_m^{all}$ increases, 
since the actual MT density $\lambda_m(t)=\lambda_m^{all}\theta(t)$ also increases, the total on-grid energy cost of all three schemes increases accordingly, 
so as to properly serve the increased MTs in the network.
It is also observed that at each value of $\lambda_m^{all}$, the total on-grid energy cost under the proposed network-level BS coordination using the Geo-DP approach is always the minimum, 
that under the cluster-based scheme is always the medium, and that under the scheme with no BS coordination is always the largest. 
Moreover, it is also worth noting that although we only show the performance of the cluster-based scheme with  two BSs in one cluster, when the number of BSs in each cluster increases,    
the performance of the cluster-based scheme is expected to  gradually approach our proposed network-level BS coordination scheme. 
However,   unlike our proposed closed-form optimal BS on/off operation policy, which has  low-complexity to coordinate all the BSs' on/off operations in the network, 
the complexity of  coordinating the increased number of  BSs in each cluster can be largely increased for the cluster-based scheme to assure the downlink transmission quality \cite{Zhou.ICTC.13}, \cite{Niu.TCOM.14}. 
As a result, our proposed network-level BS coordination scheme based on the Geo-DP approach can more efficiently solve the on-grid energy cost minimization 
problem for a large-scale network.

\section{Conclusion}

In this paper, we considered a large-scale green cellular network, where the BSs are aggregated as a microgrid with hybrid energy supplies
and an associated CES.  
We proposed a new Geo-DP approach to jointly apply stochastic geometry and the DP technique  to conduct network-level dynamic system design.
By optimally selecting  the BSs' active operation probability  as well as the amount of on-grid energy to purchase in each horizon, we studied the on-grid energy cost minimization problem.   
We first studied  the optimal BS on/off policy, which shows that the optimal BSs' active operation probability in each horizon 
is just sufficient to  assure the required downlink transmission quality with time-varying load in a large-scale network. 
We then  proposed a suboptimal on-grid energy purchase policy, where the low on-grid energy is over-purchased 
in the current horizon only when both the current storage level and future renewable energy level are both low. 
We also compared our proposed policy with the existing schemes, and showed that our proposed policy can efficiently save the on-grid energy cost in the large-scale network.  
It is also of our interest to minimize the network-level on-grid energy cost by jointly considering the uplink and downlink traffic loads in our future work.

\appendices

\section{Proof to Proposition \ref{proposition: optimal_G_T-1}}

According to (\ref{eq: J_t_G}), to find the optimal $G^*(T-1)$, we need to first find the optimal $G^*(T)$ in horizon $T$ to obtain the exact expression of 
$J_T({\bf\Lambda}(T)|{\bf\Lambda}(T-1))$ for a given system state ${\bf\Lambda}(T-1)$ in horizon $T-1$.
From (\ref{eq: B_t_G}), for a given $B(T-1)$ in horizon $T-1$, the storage level of $B(T)$ is obtained as 
\begin{equation}
 B(T)=\min\left(B(T-1)+\lambda_e(T-1)+G(T-1)-E_{min}(T-1),C \right). \label{eq: B_T_T-1}
\end{equation}
From (\ref{eq: optimal_G_T}), it is observed that  when $G(T-1)=0$, if $B(T)$ in (\ref{eq: B_T_T-1}) can be assured to be no smaller   than $E_{min}(T)-\lambda_e(T)$, 
we always have $G^*(T)=0$.  Moreover, under the assumption with $\lambda_e(T-1)<C+E_{min}(T-1)-B(T-1)$ and $C\geq \max\left(E_{min}(1),..., E_{min}(T)\right)$, 
it is easy to verify that   (\ref{eq: B_T_T-1}) can be equivalently reduced to $B(T)=B(T-1)+\lambda_e(T-1)+G(T-1)-E_{min}(T-1)$, and thus $B(T)\geq E_{min}(T)-\lambda_e(T) $  is equivalent to $B(T-1)\geq E_{min}(T)+E_{min}(T-1)-\lambda_e(T-1)-\lambda_e(T)$. 
We thus consider  two cases to find the optimal $G^*(T-1)$. One is the case with large current storage regime $B(T-1)\geq E_{min}(T)+E_{min}(T-1)-\lambda_e(T-1)-\lambda_e(T)$, 
and the other is the case with small current storage regime $B(T-1)<E_{min}(T)+E_{min}(T-1)-\lambda_e(T-1)-\lambda_e(T)$. 

First, for  case with large current storage regime, since $G^*(T-1)=0$ and thus $J_{T}({\bf\Lambda}(T)|{\bf\Lambda}(T-1))=0$, according to (\ref{eq: J_t_G}), 
we obtain $J_{T-1}({\bf\Lambda}(T-1))=\mathop{\mathrm{min}}_{ G(T-1)\geq \max(E_{min}(T-1)-B(T-1)-\lambda_e(T-1),0)}   a(T-1)G(T-1)$. As a result, 
the optimal $G^*(T-1)$ in this case is the myopic solution with $G^*(T-1)=\max(E_{min}(T-1)-B(T-1)-\lambda_e(T-1),0)$.

Next, for the case with small current storage regime, since $G^*(T)$ is not assured to be $0$ or $E_{min}(T)-B(T)-\lambda_e(T)$, $B(T)$ cannot be reduced as in the previous case without considering the storage capacity $C$. As this case is more complicated than the previous one,  based on the possible values of $G^*(T)$, we   consider and compare the following two options. 
 
\subsubsection{Option 1}
In this option, we assume $G^*(T)=0$. According to (\ref{eq: optimal_G_T}) and (\ref{eq: B_T_T-1}),   $G^*(T)=0$ holds if and only if  $\min\left(B(T-1)+\lambda_e(T-1)+G(T-1)-E_{min}(T-1),C \right)\geq E_{min}(T)-\lambda_e(T)$. Then, according to 
 (\ref{eq: J_t_G}), we solve the following optimization problem to find $G^*(T-1)$. 
 \begin{align}
\textrm{(P3)}:~~~\mathop{\mathrm{min.}}_{ G(T-1)}&~~    a(T-1)G(T-1)  \nonumber\\
\mathrm{s.t.}
& ~~ G(T-1)\geq \max(E_{min}(T-1)-B(T-1)-\lambda_e(T-1),0),  \nonumber \\
& ~~\min\left(B(T-1)+\lambda_e(T-1)+G(T-1)-E_{min}(T-1),C \right)\geq E_{min}(T)-\lambda_e(T), \label{eq: constraint_P3}
\end{align}
where  $G^*(T-1)$ is given as follows:
\begin{itemize}
 \item \emph{Case with $\lambda_e(T)\geq E_{min}(T)$}:  It is easy to verify that   (\ref{eq: constraint_P3}) always holds when  $G(T-1)=\max(E_{min}(T-1)-B(T-1)-\lambda_e(T-1),0)$ under the small current storage regime. As a result, $G^*(T-1)=\max(E_{min}(T-1)-B(T-1)-\lambda_e(T-1),0)$.
 \item \emph{Case with $\lambda_e(T)<E_{min}(T)$}: (\ref{eq: constraint_P3}) holds in this case under two scenarios. One is that $G(T-1)\geq C+E_{min}(T-1)-\lambda_e(T-1)-B(T-1)$, such that
 $\min\left(B(T-1)+\lambda_e(T-1)+G(T-1)-E_{min}(T-1),C \right)=C> E_{min}(T)-\lambda_e(T)$. The other is that $G(T-1)< C+E_{min}(T-1)-\lambda_e(T-1)-B(T-1)$, such that
 $\min\left(B(T-1)+\lambda_e(T-1)+G(T-1)-E_{min}(T-1),C \right)=B(T-1)+\lambda_e(T-1)+G(T-1)-E_{min}(T-1)$, which renders (\ref{eq: constraint_P3}) to be  equivalent to 
 $G(T-1)\geq E_{min}(T) +E_{min}(T-1)-\lambda_e(T-1)-\lambda_e(T)-B(T-1)$. It is easy to verify that the second scenario yields a smaller $G^*(T-1)=E_{min}(T) +E_{min}(T-1)-\lambda_e(T-1)-\lambda_e(T)-B(T-1)$. As a result, we obtain the cost-to-go function in this case as 
$ J_{T-1}({\bf\Lambda}(T-1))=a(T-1) \left(E_{min}(T) +E_{min}(T-1)-\lambda_e(T-1)-\lambda_e(T)-B(T-1) \right)$.

\end{itemize}

\subsubsection{Option 2}
In this option, we assume $G^*(T)=E_{min}(T)-B(T)-\lambda_e(T)$.  According to (\ref{eq: optimal_G_T}) and (\ref{eq: B_T_T-1}),   $G^*(T)=E_{min}(T)-B(T)-\lambda_e(T)$ holds if and only if  $\min\left(B(T-1)+\lambda_e(T-1)+G(T-1)-E_{min}(T-1),C \right) < E_{min}(T)-\lambda_e(T)$, which is equivalent to  $B(T-1)+\lambda_e(T-1)+G(T-1)-E_{min}(T-1) < E_{min}(T)-\lambda_e(T)$  since $C >E_{min}(T)-\lambda_e(T)$. We let $U=a(T)\left( E_{min}(T) +E_{min}(T-1)-\lambda_e(T-1)-\lambda_e(T)-B(T-1)\right)$. According to 
 (\ref{eq: J_t_G}) and (\ref{eq: B_T_T-1}),  we solve the following optimization problem to find $G^*(T-1)$ in this option. 
 \begin{align}
\textrm{(P4)}:~~~\mathop{\mathrm{min.}}_{ G(T-1)}&~~    (a(T-1)-a(T))G(T-1)+ U \nonumber\\
\mathrm{s.t.}
& ~~ G(T-1)\geq \max(E_{min}(T-1)-B(T-1)-\lambda_e(T-1),0),   \label{eq: constraint1_P4}\\
& ~~B(T-1)+\lambda_e(T-1)+G(T-1)-E_{min}(T-1) < E_{min}(T)-\lambda_e(T). \label{eq: constraint2_P4}
\end{align}
It is easy to find $G^*(T-1)$ as follows:
\begin{itemize}
 \item \emph{Case with $\lambda_e(T)\geq E_{min}(T)$}: It is observed  that if $E_{min}(T-1)-B(T-1)-\lambda_e(T-1)\geq 0$, (\ref{eq: constraint1_P4}) is reduced to 
 $G(T-1)\geq E_{min}(T-1)-B(T-1)-\lambda_e(T-1)$. However, (\ref{eq: constraint2_P4}) requires $G(T-1)<E_{min}(T)-\lambda_e(T)+E_{min}(T-1)-B(T-1)-\lambda_e(T-1)$, where the right-hand side is smaller than $E_{min}(T-1)-B(T-1)-\lambda_e(T-1)$. Similar case is also observed when $E_{min}(T-1)-B(T-1)-\lambda_e(T-1)< 0$. As a result,  
 no feasible solution of $G(T-1)$ exists in this case for Option 2, which implies that under this case the optimal $G^*(T-1)$ is obtained by using Option 1.
 \item \emph{Case with $\lambda_e(T)<E_{min}(T)$}:  Clearly, if $a(T-1)\geq a(T)$, $G^*(T-1)=\max(E_{min}(T-1)-B(T-1)-\lambda_e(T-1),0)$, with the cost-to-go function given by 
 $ J_{T-1}({\bf\Lambda}(T-1))=(a(T-1)-a(T)) \max(E_{min}(T-1)-B(T-1)-\lambda_e(T-1),0)+a(T)\left( E_{min}(T) +E_{min}(T-1)-\lambda_e(T-1)-\lambda_e(T)-B(T-1)\right)$. 
If $a(T-1)<a(T)$, $G^*(T-1)=E_{min}(T) +E_{min}(T-1)-\lambda_e(T-1)-\lambda_e(T)-B(T-1)$, with the cost-to-go function given by $ J_{T-1}({\bf\Lambda}(T-1))=a(T-1)\left(E_{min}(T) +E_{min}(T-1)-\lambda_e(T-1)-\lambda_e(T)-B(T-1) \right)$.
 \end{itemize}

Finally, by comparing the cost-to-go functions that are obtained under both options, it is easy to verify that the cost-to-go functions that are obtained under Option 2 for both cases with $a(T-1)\geq a(T)$ and $a(T-1)< a(T)$ are always no larger than that under Option 1. As a result, for the case with small current storage regime, the optimal $G^*(T-1)$ is given by that obtained under Option 2. 
Proposition \ref{proposition: optimal_G_T-1} thus follow.

\end{document}